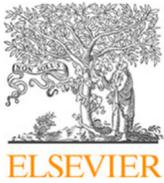

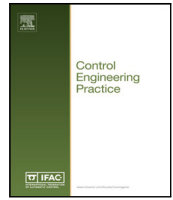

# Robust torque-observed control with safe input–output constraints for hydraulic in-wheel drive systems in mobile robots

Mehdi Heydari Shahna [ID] *, Pauli Mustalahti, Jouni Mattila

*Faculty of Engineering and Natural Sciences, Tampere University, Tampere, 33720, Finland*

## ARTICLE INFO

## ABSTRACT

Hydraulic-powered in-wheel drive (IWD) mechanisms enhance the maneuverability, traction, and maintenance efficiency of heavy-duty wheeled mobile robots (HWMRs) by enabling independent operation of each wheel. Sufficient motion in such HWMR systems relies on a multi-stage power transmission mechanism that integrates control valves, hydraulic motors, gearboxes, and, ultimately, nonlinear ground-interaction wheel dynamics on rough terrain. Deviations in each stage of these independently operated wheel systems—arising from modeling uncertainties and disturbances such as wheel slippage and uneven torque distribution on rough terrain—can disrupt motion balance between wheels and further amplify deviations. This can lead the robot to deviate from its course, oscillate, or lose traction, ultimately resulting in overall instability, which may pose a risk to the heavy-weight robot's surrounding environment. To develop a synchronous control strategy for distributed HWMR systems to mitigate such challenges in uncertain environments, this paper proposes a novel robust torque-observer-based valve control (RTOVC) framework for IWD-actuated wheels, guaranteeing robustness and uniformly exponential stability of the entire system. As a foundation for this approach, a robust torque observer network based on an adaptive barrier Lyapunov function (BLF) is designed to obtain the required wheel/motor torques, ensuring that the actual velocities of IWD-actuated wheels align with the reference values in motion dynamic frames in the presence of wheel slippages. It eliminates the closed-loop dependency on fault-prone torque or pressure sensors in hydraulic actuation mechanisms. Building on this, an additional adaptive BLF-based control network in the valve-actuated hydraulic mechanism is employed to regulate fluid flow, generating the required torque in the first network for each wheel under system uncertainties. The RTOVC framework reduces fault risks in HWMRs by constraining key input–output signals—such as valve control signals, actual wheel velocities, tracking errors, and required motor/wheel torques—within logarithmic BLFs, ensuring safe operation. A comprehensive experimental analysis on a 6,500-kg hydraulic-powered IWD-actuated HWMR operating on rough terrain, where failures may arise due to severe slipping conditions and hydraulic system uncertainties, confirms the RTOVC's robust performance compared with two other state-of-the-art control strategies.

## 1. Introduction

An in-wheel drive (IWD) system with its own motor equips each wheel of a four-wheel distributed vehicle system, enabling independent power delivery and enhanced responsiveness to various road conditions (Shao, Naghdy, & Du, 2017; Yamada, Beauduin, Fujimoto, Kanou, & Katsuyama, 2022). IWD systems eliminate traditional mechanical components of vehicles, such as differentials, by utilizing individual controllers for precise torque distribution, resulting in improved traction, reduced maintenance, and simplified motor replacements (Ghazali, 2023). This technology applies to diverse machinery and mobility systems requiring high maneuverability, making them increasingly integral to the design and functionality of off-road vehicles, such as heavy-duty wheeled mobile robots (HWMRs).

Recently, HWMRs have been extensively employed due to their operational capabilities in various industrial circumstances in which it might be inefficient, dangerous, or impossible for humans to work (Kobelski, Osinenko, & Streif, 2021; von Glehn, Gonçalves, Neto, & da Silva Fonseca, 2024). In applications such as construction, manufacturing, aerospace, and mining, many drive systems of HWMRs rely on hydraulic servomechanisms due to their power density and high torque at low speeds (Fassbender, Zakharov, & Minav, 2021). This makes them

---

# Nomenclature

The following abbreviations and symbols will be used in the forthcoming equations.

## Abbreviations

| | |
|---|---|
| **IWD** | Hydraulic in-wheel drive |
| **HWMR** | Heavy-duty wheeled mobile robots HWMR |
| **RTOVC** | Robust torque-observed valve-based control |
| **BLF** | Barrier Lyapunov function |
| **LWMR** | Light-duty wheeled mobile robot |
| **RL** | Rear left wheel (number 1) |
| **RR** | Rear right wheel (number 2) |
| **FL** | Front left wheel (number 3) |
| **FR** | Front right wheel (number 4) |
| **BAC** | Backstepping-based adaptive control |
| **DVSC** | Decentralized-valve-structure control |
| **PID** | Proportional–integral–derivative |

## Wheel Motion Dynamics

| | |
|---|---|
| $D_1$ | Slippage disturbances of the wheel |
| $F_1$ | Uncertainties of the wheel |
| $F_\omega$ | Vertical normal force of the wheel |
| $G_1$ | Known modeling term of the wheel |
| $J_\omega$ | Wheel inertia |
| $\omega_\omega$ | Actual angular velocity of the wheel |
| $\tau_\omega$ | Wheel torque |
| $a_1$ | Unknown positive coefficient of the wheel torque |
| $c_\omega$ | Damping coefficient |
| $d_\omega$ | Disturbances of the wheel |
| $f_\omega$ | Coulomb's friction of the rotation shaft of the wheel |
| $r$ | Wheel radius |
| $s_\omega$ | Slip ratio of the wheel |
| $v_\omega$ | Linear velocity of the wheel |

## Hydraulic-Powered IWD System Parameters

| | |
|---|---|
| $\tau_m$ | Hydraulic in-wheel-motor toruqe |
| $m_\omega$ | Gear ratio of each wheel |
| $p_T$ | Tank pressure or return pressure |
| $p_s$ | Source pressure provided by the hydraulic pump |
| $Q_A$ | Flows provided by the hydraulic valve at point A |
| $Q_B$ | Flows provided by the hydraulic valve at point B |
| $p_A$ | Pressure at point A |
| $p_B$ | Pressure at point B |
| $\omega_m$ | Angular velocity of hydraulic motor |
| $V_A$ | Volume of fluid per revolution at point A |
| $V_B$ | Volume of fluid per revolution at point B |
| $V_m$ | Volume of fluid per revolution |
| $\Delta p$ | Pressure difference |
| $\eta_{hm}$ | Hydromechanical inefficiencies |
| $\eta_{vol}$ | Volumetric inefficiencies |
| $\gamma$ | Effective bulk modulus of the hydraulic system |
| $Q$ | Load flow of the hydraulic system |
| $x_u$ | Normalized spool position of valves |
| $K_u$ | Flow coefficient of the valve |
| $u$ | Valve control signal |
| $\mathrm{Sat}(u)$ | Functional Constraint on valve control signal |
| $\bar{u}$ | Upper nominal valve control signal |
| $\underline{u}$ | Lower nominal valve control signal |
| $\lambda_1$ | A function between 0 and 1 |
| $B(\Delta p)$ | Nonlinear pressure-dependent function in the studied hydraulic system of the HWMR |
| $D_2$ | Unknown external disturbances in the hydraulic system |
| $F_2$ | Uncertainties in the hydraulic system |
| $\lambda_2$ | A positive function: $0 \leq \lambda_2 \leq \max(|\underline{u}|+1, |\bar{u}|+1)$ |
| $a_2$ | Unknown valve control signal coefficient |
| $A$ | Valve coefficient |
| $C$ | Hydraulic motor velocity coefficient |
| $\delta$ | A small positive term |

## RTOVC Framework Parameters

| | |
|---|---|
| $C$ | Any defined region in stability concept |
| $F_1^*$ | Unknown function: $F_1^* = F_1 - \dot{v}_d$ |
| $Y$ | Any state of a system |
| $\bar{b}$ | Positive constant |
| $\bar{c}$ | Positive constant |
| $\epsilon_1$ | Bound of the actual linear velocity: $v_\omega \leq |\epsilon_1|$ |
| $\omega_d$ | Desired angular velocity of each wheel |
| $\omega_e$ | Angular velocity error of each wheel: $\omega_e = \omega_\omega - \omega_d$ |
| $\tau_{max}$ | Bound of the motor torque: $\tau_m < |\tau_{max}|$ |
| $\zeta$ | Positive constant |
| $k_5$ | Positive constant, satisfying $k_5 \geq J_\omega = a_1^{-1}$ |
| $t$ | Time |
| $t_0$ | Initial time |
| $v_d$ | Desired linear velocity of each wheel |
| $v_e$ | Linear velocity error of each wheel: $v_e = v_\omega - v_d$ |
| $v_{max}$ | Bound of the desired linear velocity: $v_d \leq |v_{max}|$ |
| $Q_1$ | A positive notation: $Q_1 = \alpha_1^2 - v_e^2$ |
| $\alpha_1$ | Bound of velocity tracking error: $v_e < |\alpha_1|$ |
| $\beta_1$ | A positive notation: $\beta_1 = \frac{v_e}{Q_1}$ |
| $\hat{\psi}_1$ | Adaptive parameter of the first law |
| $\psi_1^*$ | Positive constant: the target of the first adaptive law |
| $\tilde{\psi}_1$ | The error of the first adaptive law: $\tilde{\psi}_1 = \hat{\psi}_1 - \psi_1^*$ |
| $k_2$ | Positive constant: Adaptive design parameter |
| $k_3$ | Positive constant: Adaptive design parameter |
| $k_4$ | Positive constant: Adaptive design parameter |

| Symbol | Description |
|---|---|
| $D_1^*$ | Unknown positive constant: $\|D_1\| \leq D_1^*$ |
| $Q_2$ | A positive notation: $Q_2 = \tau_{max}^2 - \hat{\tau}_m^2$ |
| $V_1$ | A logarithm-based BLF: $V_1 = \frac{1}{a_1}\Theta_1 + \frac{1}{k_3}\tilde{\psi}_1^2$ |
| $\Omega_{11}$ | Positive constant: $\Omega_{11} = \min[a_1 k_1, k_3 k_4]$ |
| $\Omega_{12}$ | Positive constant: $\Omega_{12} = \frac{1}{4}a_1\psi_1^{*2} + \frac{1}{2}k_4\psi_1^{*2}$ |
| $\beta_2$ | A positive notation: $\beta_2 = \frac{\hat{\tau}_m}{Q_2}$ |
| $\hat{\psi}_2$ | Adaptive parameter of the second law |
| $\hat{\tau}_\omega$ | The required wheel torque |
| $\mu_1$ | Unknown positive constant: $\|F_1^*\| \leq \mu_1 m_1$ |
| $\psi_2^*$ | Positive constant: the target of the second adaptive law |
| $\rho_1$ | Any positive constant |
| $\rho_2$ | Any positive constant |
| $\tilde{\psi}_2$ | The error of the second adaptive law: $\tilde{\psi}_2 = \hat{\psi}_2 - \psi_2^*$ |
| $k_1$ | Positive constant: torque observer design parameter |
| $k_7$ | Positive constant: Adaptive design parameter |
| $k_8$ | Positive constant: Adaptive design parameter |
| $k_9$ | Positive constant: Adaptive design parameter |
| $m_1$ | Unknown continuously positive function: $\|F_1^*\| \leq \mu_1 m_1$ |
| $D_2^*$ | Unknown positive constant: $\|D_2\| \leq D_2^*$ |
| $V_2$ | A logarithm-based BLF: $V_2 = \frac{1}{a_2}\Theta_2 + \frac{1}{k_5}\tilde{\psi}_2^2$ |
| $\Omega_{21}$ | Positive constant: $\Omega_{21} = \min[a_2 k_2, k_8 k_9]$ |
| $\Omega_{22}$ | Positive constant: $\Omega_{22} = \frac{1}{4}a_2^{-1} + \frac{1}{2}k_9\psi_2^{*2}$ |
| $\mu_2$ | Unknown positive constant: $\|F_2^*\| \leq \mu_2 m_2$ |
| $\rho_3$ | Any positive constant |
| $\rho_4$ | Any positive constant |
| $k_6$ | Positive constant: valve control design parameter |
| $m_2$ | Unknown continuously positive function: $\|F_2^*\| \leq \mu_2 m_2$ |
| $E_0$ | A non-decreasing function: $E_0 = \Theta_i^2 e^{\bar{\imath}(w-t_0)}$ |
| $E_1$ | A function: $E_1 = \sup_{w \in [t_0,t]}\left[\sum_{i=1}^2 \rho_{2i-1}^{-1}(m_i^2)e^{\bar{\imath}(w-t_0)}\right]$ |
| $Z(t)$ | A positive continuous operator: $Z(t) = \frac{0.5}{\bar{\Omega}-t}$ |
| $\Omega$ | A positive constant: $\Omega = \min[\Omega_{11}, \Omega_{21}]$ |
| $\Theta_i$ | A logarithmic function for $i=1,2$: $\Theta_i = \log(\frac{a_i^2}{Q_i})$ |
| $\alpha_2$ | Bound of the motor torque: $\alpha_2 = \tau_{max}$ |
| $\bar{E}$ | A non-decreasing function: $\bar{E} = \max[E_0, E_1]$ |
| $\bar{V}$ | A logarithm-based BLF: $\bar{V} = V_1 + V_2$ |
| $\bar{\Omega}$ | A positive constant: $\bar{\Omega} = \Omega_{12} + \Omega_{22}$ |
| $\iota$ | A positive constant: $0 \leq \iota < \Omega$ |
| $\iota^*$ | A positive constant: $\bar{\iota} < \iota^* < \Omega$ |
| $\overset{*}{Z}$ | A positive continuous operator: $\overset{*}{Z} = Z(\iota^*)$ |
| $w$ | A positive constant: $t_0 \leq w \leq t$ |
| $\bar{V}$ | A logarithm-based BLF: $\bar{V} = \sum_{j=1}^4 \bar{V}_j$ |
| $g_0$ | A defined region of stability for any IWD |
| $j$ | Number of each wheel: $j = 1, \ldots, 4$ |
| $\tau_0$ | The radius of the region $g_0$ |
| $\Omega_j$ | A positive constant: $j$th wheel's $\Omega$ |
| $\bar{V}_j$ | A logarithm-based BLF: $j$th wheel's $\bar{V}$ |
| $\bar{\Omega}_j$ | A positive constant: $j$th wheel's $\bar{\Omega}$ |
| $\bar{\Omega}$ | A positive constant: $\bar{\Omega} = \min\{\bar{\Omega}_j \mid j = 1, \ldots, 4\}$ |
| $\tilde{\Omega}$ | A positive constant: $\tilde{\Omega} = \sum_{j=1}^4 \bar{\Omega}_j$ |
| $\Theta_{i,j}$ | A logarithmic function: $j$th wheel's $\Theta_i$ with $i = 1,2$ |
| $\rho_{2i-1,j}$ | Any positive constant |
| $m_{i,j}$ | Unknown positive function: $j$th wheel's $m_i$ for $i = 1,2$ |
| $\bar{g}_0$ | A defined region of stability for the HWMR |
| $\bar{\tau}_0$ | The radius of the region $\bar{g}_0$ |

more shock-resistant and suitable for handling high loads effectively (Nonami, Barai, Irawan, & Daud, 2014).

Unfortunately, the safety and stability of HWMRs on rough terrain are compromised by wheel slippage, caused by terrain variations, load distribution imbalances, poor wheel design, dynamic forces during movement, and limitations in the robot's control systems, which may be incapable of making timely adjustments. Surprisingly, over the past twenty years, studies addressing wheel slippages among HWMRs have had relatively little focus (Teji, Zou, & Zeleke, 2023). According to our findings, the number of studies on hydraulic IWD-actuated HWMRs has notably increased in recent years, influenced by autonomous trends in heavy-duty industries, but there is still a significant shortage of robust control systems with proven stability for these high-power robots to address wheel slippages, whose development is vital to advancing the field (Hao, Quan, Qiao, Lianpeng, & Wang, 2021; Wang et al., 2023). To compensate for wheel slippages in complex off-road drive systems, tracking velocity-based control has minimal computational complexity (Teji et al., 2023; Wang, Gao, Wang, Wang, & Wang, 2021). However, the wheel torque in this control strategy tends to be generated excessively (Teji et al., 2023; Wang et al., 2021), leading to soil terrain failure (Teji et al., 2023) and a harmful increase in the generation of other key signals, such as valve control, tracking error, and actual velocity of each hydraulic IWD. This excessive torque may pose threats to adjoining facilities, compromise the system's mobility, and lead to failures. Therefore, it is essential to impose safe constraints on key signals to reduce the risk of system instability and faults in hydraulic-powered, IWD-actuated HWMRs operating in uncertain and time-varying conditions.

In addition, controlling hydraulic servomechanisms conventionally requires numerous fault-prone sensors. For instance, feedback torque/pressure information, either directly by using a motor torque/force sensor or indirectly through calculations based on pressure sensors for opening valves is required to generate sufficient torque efforts applied to in-wheel motors (Lee, Kim, & Chung, 2016). However, incorporating such sensors not only increases the cost of designing an HWMR but also renders the control system more dependent on sensors susceptible to faults in a harsh environment with heavy loads (Vo, Dao, Ahn, et al., 2021; Yin, Xu, Fan, & Sun, 2024). For instance, to eliminate the necessity for pressure sensors, Estrada, Ruderman, Texis-Loaiza, and Fridman (2025) proposed a third-order sliding mode control aligned with a finite-time output derivative observer for robust trajectory tracking of a hydraulic cylinder, relying on measuring the position of the cylinder rod in the presence of friction, other nonlinearities, and disturbances. Similarly, to eliminate the necessity of velocity sensors, Estrada, Ruderman, and Fridman (2024) provided a novel integral sliding surface for noisy hydraulic systems to track a sufficiently smooth reference.

To address robustness against such challenges for HWMRs in uncertain and time-varying operating conditions, research studies documented in Chen et al. (2023), Galati, Mantriota, and Reina (2022),

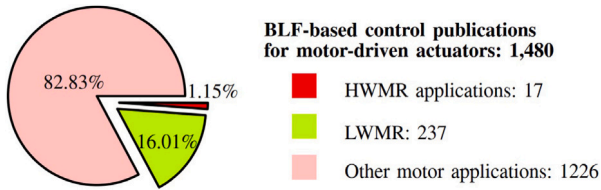

**Fig. 1.** Distribution of BLF-based control research for motor-driven actuators with any power type Since 2020.

Nagariya and Saripalli (2020), Wang, Fader, and Marshall (2023) consistently emphasized on tracking control, managing modeling uncertainties, and mitigating external disturbances. In addition, further advancements reported by Gauthier-Clerc, Hill, Laneurit, Lenain, and Lucet (2021), Knaup, Okamoto, and Tsiotras (2023), Zhang et al. (2024) took steps forward by addressing safety constraints in certain aspects, thereby broadening the scope and applicability of these control strategies. However, further advancements are still required to design a robust, highly stable, safe and robust control system that decreases the closed-loop dependency associated with fault-prone sensors specifically for valve-actuated IWD-actuated HWMRs.

It is beyond doubt that key control signals of IWD-actuated HWMRs must safely be constrained within predefined safety limits to prevent system faults, minimize wheel slippage, and enhance operational safety in environments characterized by highly uncertain and unstructured conditions, but incorporating these safety constraints into the control system, while simultaneously ensuring both robustness and stability, remains the primary challenge. According to ISO/IEC TR 5469 (International Organization for Standardization, 2024), a control design using barrier function approaches can guarantee the system operation remains within a safe region. For example, a barrier Lyapunov function (BLF) control extends the traditional Lyapunov concept, mainly focused on the stability or convergence of system states, by incorporating constraints that the system should never violate (Liu et al., 2023; Yu, Zhou, Sun, Sang, & Zhang, 2024). This capability has led to more research studies on BLF-based control for various applications. Based on our findings, Fig. 1 shows the number of BLF-based control studies for various applications utilizing motor-driven actuators, including HWMRs, light-duty wheeled mobile robots (LWMRs), and other applications since 2020. These data demonstrate that the capabilities of BLF in HWMR control have not yet been fully realized. All seventeen BLF-based control publications cannot apply to our case study, which involves highly nonlinear hydraulic-powered IWD-actuated HWMR systems prone to sensor faults and severe wheel slippage.

This paper proposes a novel robust torque-observer-based valve control (RTOVC) framework for independently IWD-actuated HWMR platforms. To begin with, a new torque observer network is designed based on the adaptive BLF to obtain the necessary torque to align the wheel velocity with the reference one and effectively reject the imposed disturbances due to rough terrains, wheel slippage, and other wheel effects. Concurrently, the obtained torque, as the reference torque for the hydraulic system, is fed into another adaptive BLF in the hydraulic actuation mechanism, which regulates valve control signals to modulate the hydraulic pressure to either increase or decrease the force exerted by the hydraulic fluid to achieve the required torque for each actuator. As a summary, this work provides the following new findings in the field:

(1) to eliminate the closed-loop dependency on fault-prone torque (or hydraulic pressure) sensors, a novel robust wheel/motor torque observer network is introduced in the RTVOC framework. This network determines the reference torque in the hydraulic actuation mechanisms based on the required wheel motion under uncertainties and disturbances;

(2) to address the issue of excessive torque generation, a common problem in velocity-based control systems (Teji et al., 2023; Wang et al., 2021), particularly during wheel slippage, the hydraulic motor torque is safely constrained within a logarithmic BLF framework;

(3) to reduce the risk of system faults occurring in IWD-actuated HWMRs, other key input–output signals—such as the actual velocity and tracking errors of IWD-actuated wheel and valve control signals—are constrained within another logarithmic BLF framework;

(4) the RTOVC-applied IWD-actuated wheels contribute to the robustness and stability of the entire distributed HWMR system into a specified stable region with an exponential convergence rate. The radius of the stable region adaptively depends on the intensity of slippage, load, and rough terrain effects.

The remainder of this paper is organized as follows: in Section 2, hydraulic IWDs are modeled. First, the motion dynamics of the wheel are discussed, including a mathematical description of wheel slippage. Second, the hydraulic motor formulation is developed by considering the valve model, providing the dynamic formulation of the hydraulic actuation mechanisms under safely defined constraints of the valve controls. Section 3 presents the step-by-step design of the RTOVC, ensuring the uniformly exponential stability of the RTOVC-applied hydraulic servomechanism for each wheel of an HWMR and, consequently, the stability of the entire system. Finally, Section 4 demonstrates the validity of the RTOVC strategy by presenting a comprehensive experiment on a 6500-kg hydraulic-powered IWD-actuated HWMR across two challenging terrains: (1) snow-covered gravel terrain with soft soil and (2) steep ice-covered stony terrain.

## 2. Modeling hydraulic IWDs

Normally, the tractive force generated by each wheel on a surface propels an HWMR forward by overcoming various resistances. This force is directly related to the torque received from the in-wheel driving gear (hydraulic motor) $\tau_m$ and is converted into the driven gear (wheel) $\tau_\omega$. The linear dynamics of each driving wheel under slippage can be provided, as (Liao, Chen, & Yao, 2018; Petrović & Mattila, 2022):

$$\dot{v}_\omega = \frac{r}{J_\omega}\left(\tau_\omega - rF_\omega - c_\omega \omega_\omega - f_\omega + d_\omega\right) - \dot{s}_\omega \tag{1}$$

where $v_\omega$ and $\omega_\omega$ are each wheel's linear and angular velocities, respectively. The damping coefficient is represented as $c_\omega$, and the Coulomb's friction of the rotation shaft as $f_\omega(\omega_\omega)$. $d_\omega$ represents the disturbance, $F_\omega$ is the vertical normal force of each wheel, $J_\omega$ is the wheel inertia, and $r$ is the wheel radius. The wheel slip for IWDs is defined as an unknown function $s_\omega$. We simplify them, as follows:

$$\dot{v}_\omega = a_1 \tau_\omega + G_1(\omega_\omega, t) + F_1(\omega_\omega, t) \tag{2}$$

where $a_1 = r/J_\omega$ is a bounded coefficient. Although $a_1$ is a well-known system parameter from a machine construction perspective, we assume its value is unknown for control design, as the goal is to design a generic control system that operates independently of the wheel torque coefficient in motion dynamics. This reduced dependency on modeling terms may simplify control implementation. $G_1$ is assumed to be the known modeling term, $F_1$ represents the unknown modeling error due to state-variant uncertainties and time-variant external disturbances.

**Assumption 1.** Assume the control gain $a_1$ is positive, but its finite value is unknown (in contrast to Oliveira, Peixoto, & Hsu, April 2015). The IWD modeling error $F_1(\omega_\omega, t)$ are unknown but bounded (in contrast to Huang and Liu (January 2019), where partial knowledge of the uncertainty function or its corresponding bounding function is predefined). This implies that for all values of $\omega_\omega$ within its domain and for all times $t$, $|F_1(\omega_\omega, t)|$ can always be bounded above by a positive function and cannot grow infinitely large.

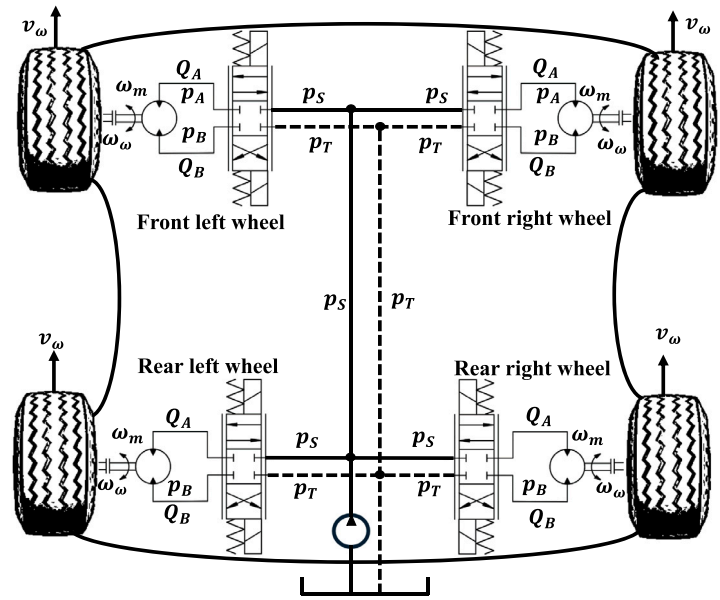

**Fig. 2.** A hydraulic IWD-actuated HWMR with four independent wheels.

The valve-based servomechanism based on Hulttinen and Mattila (2021), Jelali and Kroll (2012) for each wheel can be modeled. An HWMR actuated by four independently valve-controlled hydraulic motors is shown in Fig. 2. Note that $p_T$ is the tank pressure or return pressure, $p_s$ is the source pressure provided by one hydraulic pump (supplying energy to the four-wheel systems), $Q_A$ and $Q_B$ are the flows provided by the hydraulic control valves, $p_A$ and $p_B$ are pressures at different points in the hydraulic system, $\omega_m$ is the angular velocity of the hydraulic motor.

Each in-wheel motor torque can be generated as the pressure difference between the high-pressure line and low-pressure line, as (Hulttinen & Mattila, 2021):

$$\tau_m = \Delta p \frac{V_m}{2\pi} \eta_{hm} \tag{3}$$

where we assume both line volumes are equal to the motor displacement $V_A = V_B = V_m$ (volume of fluid per revolution) and $\Delta p = p_A - p_B$. We assume $\eta_{hm}$ demonstrates unknown hydromechanical inefficiencies. The differential pressure dynamic of the hydraulic motor based on the flow can be provided as (Hulttinen & Mattila, 2021; Inderelst & Murrenhoff, 2014):

$$\Delta \dot{p} = \frac{\gamma}{V_m} \left( Q - \omega_m \frac{V_m}{\pi} \eta_{vol} \right) \tag{4}$$

We assume volumetric inefficiencies $\eta_{vol}$ such as internal leakage, may exist and are unknown. $\gamma$ is the effective bulk modulus of the system, and $Q = Q_A - Q_B$ is the load flow. From Hulttinen and Mattila (2021), Inderelst and Murrenhoff (2014), the flow-pressure dynamics of motors can be provided based on Eqs. (3) and (4) as:

$$Q = \dot{\tau}_m \frac{2\pi}{\gamma \eta_{hm}} + \omega_m \frac{\gamma V_m}{\pi} \eta_{vol} \tag{5}$$

This equation represents the required motor flow, where the first term accounts for the effects of fluid compression on torque variations, while the second term captures the influence of motor rotation. This required volumetric flow is regulated by the valve control signal $u$. The studied valve model in terms of required motor flow can be expressed as (Hulttinen & Mattila, 2021):

$$Q = K_u u \sqrt{2 \left( p_S - \text{sign} \left( x_u \right) \Delta p + \delta \right)} \tag{6}$$

which links the normal spool position $x_u$ to the flow rate. $K_u$ is the flow coefficient of the valve. In addition:

$$\text{sign}(x_u) = \begin{cases} -1 & \text{if } x_u < 0, \\ 0 & \text{if } x_u = 0, \\ 1 & \text{if } x_u > 0. \end{cases} \tag{7}$$

Since torque generation is dependent on flow rate, the valve acts as the actuator that translates control commands into real-time torque adjustments by managing fluid supply to the motor. Note that $\delta$ in Eq. (6) is a small positive term to prevent numerical singularity in the equation. Physically, this could correspond to a scenario where the downstream pressure is zero (e.g., due to cavitation or extreme flow conditions), or the valve is fully open, allowing unrestricted flow, which is unrealistic due to practical flow limitations. In addition, implementing physical constraints in the control signals is provided later in Eq. (11) to help avoid extreme pressure drops. Under common modeling assumptions, let us define two constants:

$$A = \frac{\gamma \eta_{hm} K_u}{2\pi}, \quad C = \frac{\gamma V_m}{\pi} \eta_{vol} \tag{8}$$

In addition, we define a nonlinear function $B(.)$ which is pressure-dependent, as:

$$B(\Delta p) = \sqrt{2 \left( p_S - \text{sign} \left( x_u \right) \Delta p + \delta \right)} \tag{9}$$

where we assume it is bounded by an unknown constant and does not grow without bound. Now, we insert the studied valve model provided in Eq. (6) into the differential motor torque in Eq. (5). Finally, from (5) and (6) and building on Pasolli and Ruderman (2020), the hydraulic actuation mechanism can be presented in the form of a hydraulic system linearization, as:

$$\dot{\tau}_m = A \ B(\Delta p) \ u - C \ \omega_m \tag{10}$$

Eq. (10) is the dynamic model of the studied hydraulic system, including the valve and hydraulic-motor models used to actuate each wheel of the HWMR. In addition, valve control signal constraints for each hydraulic servomechanism can be defined as:

$$\text{Sat}(u) = \begin{cases} \bar{u}, & u \geq \bar{u} \\ u & \underline{u} \leq u \leq \bar{u} \\ \underline{u} & u \leq \underline{u} \end{cases} . \tag{11}$$

where $\bar{u}$ and $\underline{u}$ specify the upper and lower nominal valve control signals, respectively. Consequently, we define:

$$Sat(u) = \lambda_1 \ u + \lambda_2 \tag{12}$$

where

$$\lambda_1 = \begin{cases} \frac{1}{|u|+1}, & u \geq \bar{u} \ \text{or} \ u \leq \underline{u} \\ 1 & \underline{u} < u < \bar{u} \end{cases} \tag{13}$$

and

$$\lambda_2 = \begin{cases} \bar{u} - \frac{u}{|u|+1}, & u \geq \bar{u} \\ 0 & \underline{u} < u < \bar{u} \\ \underline{u} - \frac{u}{|u|+1} & u \leq \underline{u} \end{cases} \cdot \tag{14}$$

Eqs. (12), (13), and (14) imply Eq. (11). Note $0 \leq \lambda_2 \leq \max(|\underline{u}|+1, |\bar{u}|+1)$ and $0 \leq \lambda_1 \leq 1$. Thus, by considering (11), we will have (10) as the following form:

$$\dot{\tau}_m = A \ B(\Delta p) \ Sat(u) - C \ \omega_m \tag{15}$$

From (12) and (15):

$$\dot{\tau}_m = A \ B(\Delta p) \ \lambda_1 u + A \ B(\Delta p) \ \lambda_2 - C \ \omega_m \tag{16}$$

Similar to (2), let us to simplify (16), as follows:

$$\dot{\tau}_m = a_2 \ u + F_2(\omega_m, \Delta p, t) \tag{17}$$

where $a_2 = A \ B(\Delta p) \ \lambda_1$ is a bounded time-varying coefficient. Since the aim is to design a generic control system that operates independently of the control valve coefficient in hydraulic mechanisms, we assume the value of $a_2$ is unknown for control design. This reduced dependency on modeling terms can simplify control implementation. $F_2$ represents the unknown modeling error due to state-variant and time-variant uncertainties in the hydraulic system.

**Assumption 2.** Similar to Assumption 1, assume the control valve signal gain $a_2$ is positive, but its finite value is unknown for the control design. The uncertainty and disturbance of the hydraulic system $F_2(\omega_m, \Delta p, t)$ are unknown but bounded. This implies that for all values of $\omega_m$ and $\Delta p$ within their domain and for all times $t$, $|F_2(\omega_m, \Delta p, t)|$ can always be bounded above by a positive function and cannot grow infinitely large. Thus, let us introduce $\mu_2 \in \mathbb{R}^+$ as an unknown positive constant, and $m_2 : \mathbb{R} \to \mathbb{R}^+$ as a continuously bounded function with strictly positive values. They assign the upper bound of $F_2$, as:

$$|F_2| \leq \mu_2 \ m_2(t) \tag{18}$$

## 3. RTOVC framework design

The tracking error of each wheel is introduced, as follows:

$$v_e(t) = v_{\omega}(t) - v_d(t), \quad \omega_e(t) = \frac{v_e(t)}{r} = \omega_{\omega}(t) - \omega_d(t) \tag{19}$$

where $\omega_e$ and $v_e$ are the tracking angular and linear velocity error of the wheel between the actual angular/linear ($\omega_{\omega}$ or $v_{\omega}$) and reference ($\omega_d$ or $v_d$) velocity of the wheel. By considering Eq. (2), and making a derivative of (19) for linear velocity error, we will have:

$$\dot{v}_e(t) = a_1\tau_m(t) + G_1(\omega_{\omega}, t) + F_1(\omega_{\omega}, t) - \dot{v}_d(t) \tag{20}$$

Let us define $F_1^*(\omega_{\omega}, t, \dot{v}_d) = F_1(\omega_{\omega}, t) - \dot{v}_d(t)$. Thus:

$$\dot{v}_e(t) = a_1\tau_m(t) + G_1(\omega_{\omega}, t) + F_1^*(\omega_{\omega}, t, \dot{v}_d)(t) \tag{21}$$

**Assumption 3.** Assume $\dot{v}_d$ exists and is bounded. Based on Assumption 1 and similar to Assumption 2, we can assume $F_1^*$ is also bounded above by a positive function as

$$|F_1^*(\omega_{\omega}, t, \dot{v}_d)| \leq \mu_1 \ m_1(t) \tag{22}$$

where $m_1 : \mathbb{R} \to \mathbb{R}^+$ is a continuously bounded function with strictly positive values, and $\mu_1 \in \mathbb{R}^+$ is an unknown positive constant.

Assume the reference linear velocity is bounded, such that $|v_d| \leq v_{max}$ where $v_{max}$ is a positive constant. After introducing the valve control signal constraint (system input constraint) in Eq. (11), let us apply safety-defined wheel constraints on actual linear velocity, tracking error, and motor/wheel torque, as follows:

$$|v_{\omega}| < \epsilon_1, \quad |v_e| < \alpha_1, \quad \tau_m < \tau_{max} \tag{23}$$

where $\tau_{max}$ is a positive constant that can be adjusted to lower values on different terrains to address the issue of excessive motor torque generation, a common problem in velocity-based control systems, particularly during wheel slippage. $\epsilon_1$ and $\alpha_1$ are also positive constants to satisfy these conditions: $v_{max} < \epsilon_1$, $\alpha_1 = \epsilon_1 - v_{max}$. Let us define an adaptive law $\dot{\psi}_1$, based on (Shahna & Abedi, 2021):

$$\dot{\psi}_1 = -k_3 k_4 \dot{\psi}_1 + \frac{1}{2} k_2 k_3 \beta_1^2 \tag{24}$$

$k_2, k_3, k_4$ are positive constants. $\beta_1$ is a positive notation and is defined as:

$$\beta_1 = \frac{v_e}{\alpha_1^2 - v_e^2} \tag{25}$$

As observed in (25), if $|v_e(t_0)| < \alpha_1(t_0)$, the velocity error must satisfy $|v_e| < \alpha_1$. Increasing $v_e^2$ and reaching the predefined tracking error bound $\alpha_1^2$ leads to singularities in the system. For instance, in many programming platforms, execution can stop due to a warning, such as 'Infinity or NaN value encountered,' when the computed value exceeds the system's numerical limits. This characteristic is useful to ensure that the tracking error variable does not exceed bound $\alpha_1$, which could be unsafe or undesirable in a control context.

**Remark 1.** As heavy-duty IWD-actuated HWMRs must adhere to high safety standards, a primary objective of this paper is to halt operation only when the system approaches unsafe performance, utilizing barrier functions. Following ISO/IEC TR 5469 (International Organization for Standardization, 2024), and based on the BLF concept, the logarithmic or fractional barrier function can be designed, tending to infinity as the system state approaches a constraint boundary. These barrier functions operate similarly to an emergency stop button when unsafe performance, but do so automatically and inherently within the communication protocol, without requiring continuous operator supervision.

Let us propose the required wheel torque based on velocity tracking errors ($\hat{\tau}_{\omega}$) as

$$\hat{\tau}_{\omega} = -\frac{1}{2}(k_1 v_e + k_2 \hat{\psi}_1 \beta_1) - k_5 \beta_1 G_1^2 \tag{26}$$

$k_1$ and $k_5$ are positive constants. From (26), we can obtain the required motor torque as:

$$\hat{\tau}_m = \frac{\hat{\tau}_{\omega}}{m_{\omega}} \tag{27}$$

where $m_{\omega}$ is the gear ratio between the motor and the driven wheel. Similar to (24), let us introduce another adaptive law $\dot{\psi}_2$, as:

$$\dot{\psi}_2 = -k_8 k_9 \dot{\psi}_2 + \frac{1}{2} k_7 k_8 \beta_2^2 \tag{28}$$

where $k_7, k_8$, and $k_9$ are positive constants. $\beta_2$ is a positive notation and is defined, as:

$$\beta_2 = \frac{\hat{\tau}_m}{\tau_{max}^2 - \hat{\tau}_m^2} \tag{29}$$

As observed in (29), if $|\hat{\tau}_m(t_0)| < \hat{\tau}_{max}(t_0)$, the motor torque must satisfy $|\hat{\tau}_m| < \tau_{max}$. Increasing $\hat{\tau}_m^2$ and approaching $\tau_{max}^2$ leads to singularities in the system. For instance, in many programming platforms, execution may stop due to a warning, such as 'Infinity or NaN value encountered,' when the computed value exceeds the system's numerical limits. It can prevent the issue of excessive motor torque generation, a

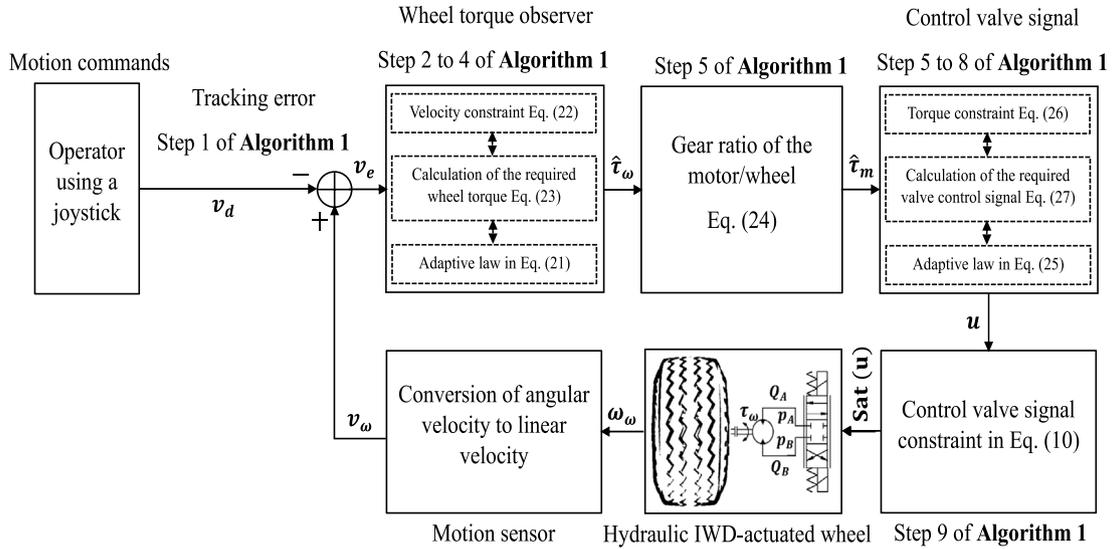

**Fig. 3.** Integration of the proposed RTOVC for each hydraulic IWD-actuated wheel in the HWMR, considering safety constraints and operating independently of torque feedback.

common problem in velocity-based control systems, particularly during wheel slippage. Now, let us suggest the valve signal control, as follows:

$$u = -\frac{1}{2}(k_6\hat{\tau}_m + k_7\hat{\psi}_2\beta_2) \tag{30}$$

$k_6$ is a positive constant. Since the entire hydraulic system model is assumed to be unknown, the proposed control in (30) does not include a modeling term $G_2$ as seen in (26). The step-by-step integration of the individual RTOVC into each valve-based IWD of HWMRs is depicted in Algorithm 1 and Fig. 3.

---

**Algorithm 1** Step-wise guidance of the RTOVC framework

---

**Input:** Measured velocity $v_\omega$ and reference velocity $v_d$.
**Output:** Constrained control signal Sat($u$).
1   Obtain the velocity tracking error in each wheel $v_e$: Eq. (19);
2   Calculate the velocity safety constraint notation $\beta_1$: Eq. (25);
3   Obtain the first adaptive parameter $\hat{\psi}_1$: Eq. (24);
4   Obtain the required torque $\hat{\tau}_\omega$: Eq. (26);
5   Calculate required hydraulic motor torque $\hat{\tau}_m$: Eq. (27);
6   Calculate the torque safety constraint notation $\beta_2$: Eq. (29);
7   Obtain the second adaptive parameter $\hat{\psi}_2$: Eq. (28);
8   Obtain control valve signal $u$: Eq. (30);
9   Constrain the control valve signal Sat($u$) Eq. (12);
10  Back to step 1.

---

Fig. 3 shows that the user/operator via a joystick dictates motion commands. Based on this, each hydraulic-powered IWD-actuated wheel system in the studied HWMR employs a two-stage control framework integrating the robust torque observer and the valve control to ensure precise motion tracking and system stability. Observing the closed-loop motion tracking error, the first control stage obtains the required wheel motor/wheel torque. The gear ratio governs the required torque and speed transformation between the motor and the wheel. The proposed control valve signal in the second stage ensures sufficient actuation control effort based on the required motor torque, while adhering to the valve signal constraint. The system leverages adaptive laws to compensate for uncertainties and variations in both wheel motion dynamics and the hydraulic IWD mechanism, ensuring robustness under changing operating conditions.

**Theorem 1.** *Under Assumptions 1 and 2, the proposed RTOVC framework guarantees the uniformly exponential stability of the hydraulic IWD mechanism described in Eqs. (1) and (10) in tracking the reference wheel motion, despite state- and time-variant uncertainties and safely predefined constraints.*

**Proof.** See Appendix A.

**Theorem 2.** *The stability of the RTOVC framework for each hydraulic-powered IWD mechanism in Theorem 1 ensures the uniformly exponential stability of the entire HWMR system with four independently operated IWD mechanisms*

**Proof.** See Appendix B.

## 4. Experimental results and analysis

### 4.1. Experimental setup

This section investigates the onboard implementation of the RTOVC applied to four independently high-bandwidth valve-based hydraulic-powered IWD-actuated wheels with the same diameter ($r = 0.854$ m) of a 6650-kg HWMR. The gear ratio of the reducer transmission for the wheel and in-wheel motor is $m_\omega = 17.7$. This 6650-kg hydraulic HWMR is the Haulotte HA16 RTJ Pro available in the Innovative Hydraulics and Automation group lab in the Tampere University provided by industrial sponsors for research and development purposes. The default control system used in the studied HWMR was an open-loop control, where the operator commanded the system via a joystick. The state-of-the-art closed-loop control for this system, provided in Hulttinen and Mattila (2021), was a model-based PD control that addressed neither robustness nor provided stability proof. The instrumentation and hardware of this setup are also provided in Fig. 4 and Table 1. In addition, the angular velocity of each wheel $\omega_\omega$ was measured from an in-wheel electromagnetic speed sensor, which brings reliable navigation capabilities to each hydraulic-powered IWD-actuated wheel control of the studied HWMR. The speed sensor detects the shaft speed and the direction of rotation, being mounted to the encover of the Danfoss motor, and senses the speed from a magnet that is rotating inside the motor. The linear velocity of each wheel was calculated by $v_\omega = r\omega_\omega$ The communication between the case study and the computer is provided in Fig. 5. As discussed, slippage and low friction between each wheel and the ground may introduce intense and unpredictable disturbances to the other wheels, challenging the platform's stability. Thus, when heavy-duty wheels rotate on rough or slippery terrain, the heavy-weight HWMR may lose control and stability during motion, potentially resulting in damage (Deniz, Jorquera, Torres-Torriti, & Cheein, 2024; Nonami et al., 2014).

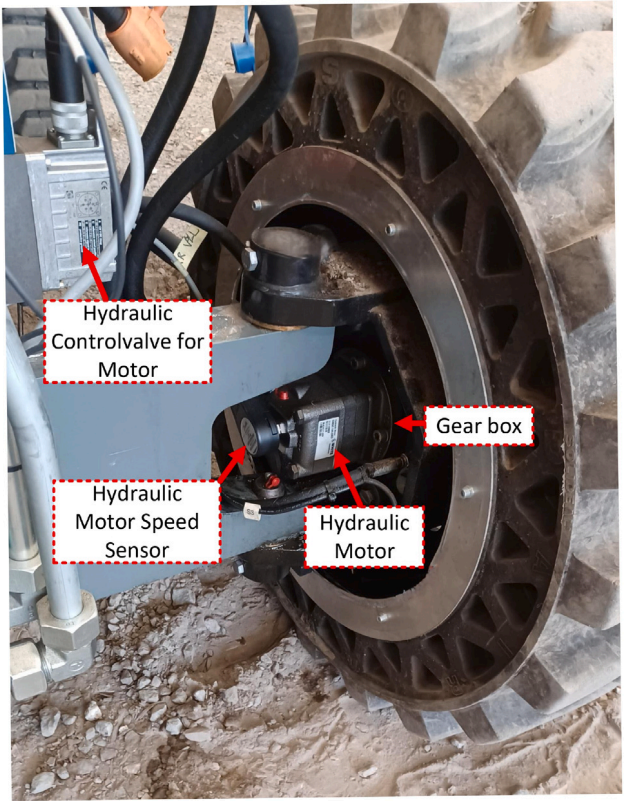

**Fig. 4.** Experimental setup of the studied hydraulic-powered IWD wheel.

**Table 1**
Instrumentation and hardware configuration of four hydraulic-powered IWD-actuated wheels of the studied HWMR system.

| Component | Description |
|---|---|
| Kubota Diesel Engine | 26.5 kW @ 3000 rpm |
| Bosch Rexroth Pump | 63 l/min |
| Danfoss OMSS Motors | 100 cm³/rev |
| Bosch Rexroth valves | 40 l/min@$\Delta p$ = 3.5MPa |
| IFM PA3521 transducers | sensor range: 25 MPa |
| Danfoss EMD Speed Sensor | 0–2500 rpm |
| Beckhoff IPC CX2030 | 1000 Hz sample rate |

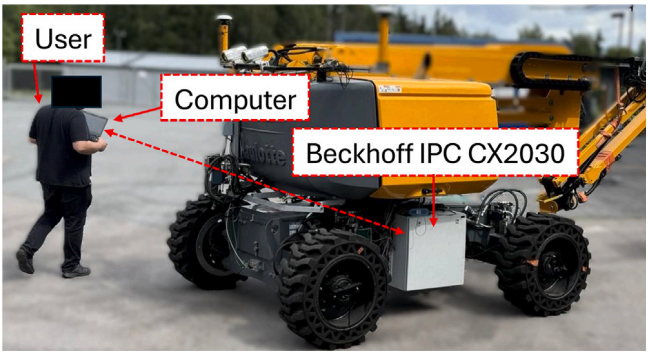

**Fig. 5.** Communication of the setup including sensory information and control commands.

To define realistic challenging scenarios for assessing the robust performance of RTOVC applied to four hydraulic-powered IWD-actuated wheels of the HWMR, the following tasks were considered:

- *Experiment 1:* The studied HWMR moves on a snow-covered gravel terrain with soft soil and uneven surfaces. In this scenario, the

IWDs are particularly susceptible to slippage due to the non-distributed heavy weight of the HWMR in the spot of the interaction between the wheel and the soft snowy surface. This scenario implicitly represents realistic off-road conditions that a IWD-actuated HWMR may experience.

- *Experiment 2:* The studied hydraulic-powered IWD-actuated HWMR moves on a steep ice-covered stony terrain. In this scenario, the increased gravitational resistance (during upward movement) and strain on the brakes (during backward movement) can cause the 6650-kg HWMR to skid or tip. In addition, the slippery icy surface complicates this situation.

Both scenarios are presented as image sequences (from left to right hand) in Fig. 6. Here, we abbreviate four hydraulic-powered IWD-actuated wheels of the HWMRs, as follows: front/left (FL), front/right (FR), rear/left (RL), and rear/right (RR) wheels. To investigate the capability of the motor torque observer (26) in the RTOVC framework, we did not have any information on the pressure or torque sensors of the wheels in the closed-loop control. The input to the RTOVC was the error between the in-wheel electromagnetic speed sensor-based actual linear velocities of the wheels (calculated from the motor velocities, gear ratios, and wheel radius) and the reference linear velocity generated by a joystick. The output consisted of four constrained control valve signals (see Fig. 7). In addition, the safety-defined constraints in the RTOVC framework were defined, based on (23), in both experiments, as:

$$|v_o| < \epsilon_1 = 0.5 \text{ m/s}, \qquad |\tau_m| < \tau_{max} = 290 \text{ N.m}$$
$$|\text{Sat}(u)| \le \bar{u} = u_{max} = 0.44, \qquad |v_e| < \alpha_1 = 0.5 - v_{max} \tag{31}$$

$v_{max} = 0.25$ m/s is the maximum value of the reference wheel velocity ($v_d$) generated by the joystick command, adjusted in the joystick section built in the Beckhoff before each operational task. Furthermore, the maximum and minimum generated valve control signals were constrained as $u_{max}$ and $-u_{max}$. The control parameters we used in this study are, as follows: $k_1 = k_6 = 3$, $k_2 = k_3 = k_4 = k_7 = k_8 = k_9 = 1$, and $k_5 = 100$. Interestingly, except for $k_1$ and $k_6$, the control performance of RTOVC was not visibly dependent on the other control parameters; being positive was sufficient for other constant parameters. A guide on parameter tuning of the proposed RTOVC is provided in Appendix C.

### 4.2. Experiment 1: snow-covered gravel terrain

The large-scale image of the IWD-actuated HWMR on the soft and rough terrain is shown in Fig. 8.

During this execution, all wheels experienced slippages; in particular, the RL and FR wheels exhibited severe slippages during backward movement over a highly uneven spot with an extremely soft ground layer, which occurred between 80 and 90 seconds after beginning the task. Despite this, the wheels' velocity tracking is shown in Fig. 9, indicating that all wheels, utilizing the RTOVC, tracked the joystick-generated velocity (depicted in black) under severe slippage, ensuring the robustness of control without torque feedback sensing. The safety-defined constraints on the actual velocity $|v_o| < 0.5$ m/s and velocity tracking error $|v_e| < 0.25$ m/s which were defined in Eq. (31) were also met. As observed, the actual velocities of the IWD-actuated wheels varied due to the uneven and non-smooth terrain, while still tracking the reference command. The large-scale image in Fig. 9 also illustrates the quick response of the RTOVC to changes in the velocities of the IWD-actuated wheels.

The actual tracking error signals are shown in Fig. 10, which more clearly illustrates the control performance across all four wheels. The figure clearly demonstrates that all four wheels consistently suffered from slippage, with the RL and FR wheels experiencing more severe slippage. This validates that the control performance remains within the safety-defined error constraints while maintaining robustness in the presence of severe disturbances caused by rough terrain on the studied 6650-kg HWMR. The control valve signals of the RTOVC for this scenario are presented in Fig. 11. It generates sufficient valve signals to produce the required torques, consequently tracking the commanded velocity from the joystick. As observed during this scenario, the uneven and non-smooth terrain caused the control signals to be generated excessively, leading to system instability and soil failure. However, the safety-defined constraints on all four valve signals as (31), $|sat(u)| \leq 0.44$, were adhered to. This behavior is further corroborated by Fig. 12, which illustrates the required torques. This demonstrates the required four motor torques obtained from Eq. (26) to generate sufficient wheel motion for tracking the commanded joystick reference motion while maintaining the stability of the HWMR under slippage on snowy terrain. Similar to the constrained control valve signals and IWD velocities, the required motor torque never exceeded the safety-defined constraints, $|\tau_m| < 290$, as shown in Fig. 12, even under slippages.

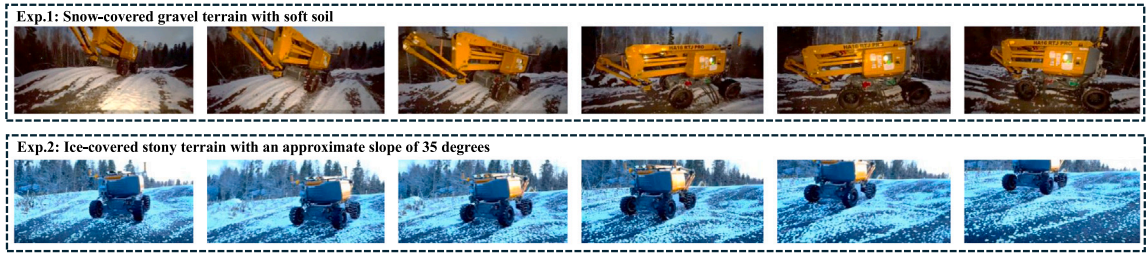

**Fig. 6.** Experimental scenarios for the studied hydraulic-powered IWD-actuated HWMR are as follows: (1) snow-covered gravel terrain with soft soil and (2) ice-covered stony terrain with an approximate slope of 35 degrees. The relevant videos are available at: https://youtu.be/4vSnfQ9PQAA and https://youtu.be/AWnTkHOndLU.

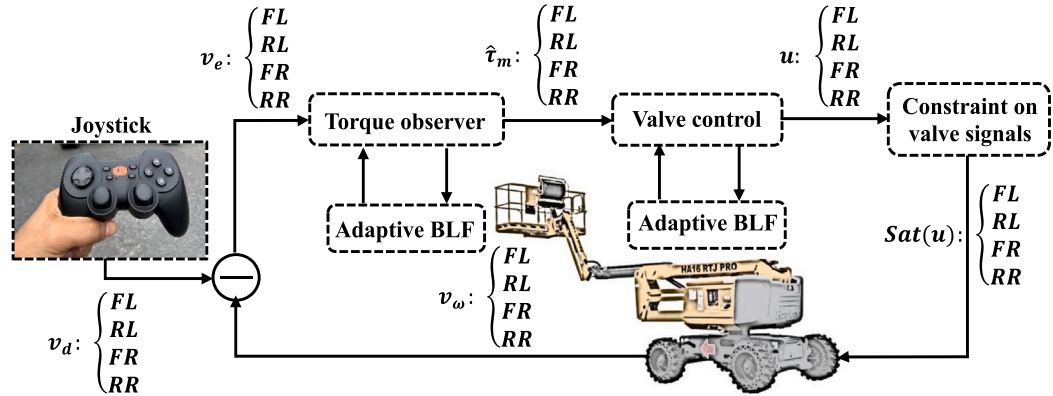

**Fig. 7.** The RTOVC-applied hydraulic-powered IWD-actuated HWMR in practice.

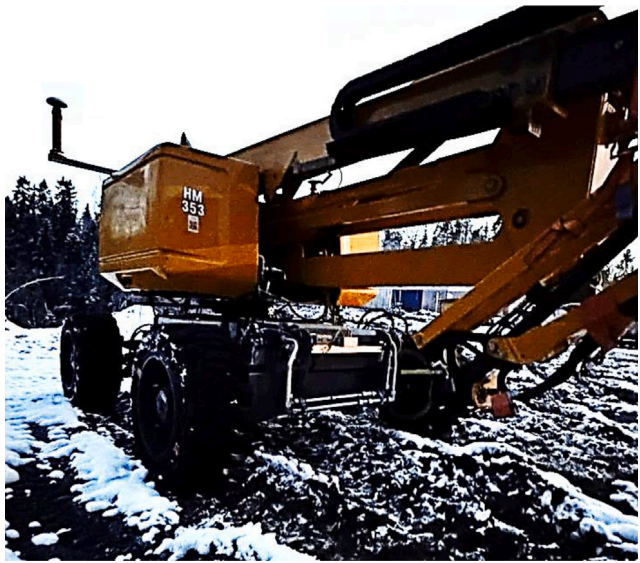

**Fig. 8.** In Experiment 1, the performance of the studied HWMR was evaluated on a rough gravel terrain with a soft nature, covered by a layer of snow. The relevant video is available at: https://youtu.be/4vSnfQ9PQAA.

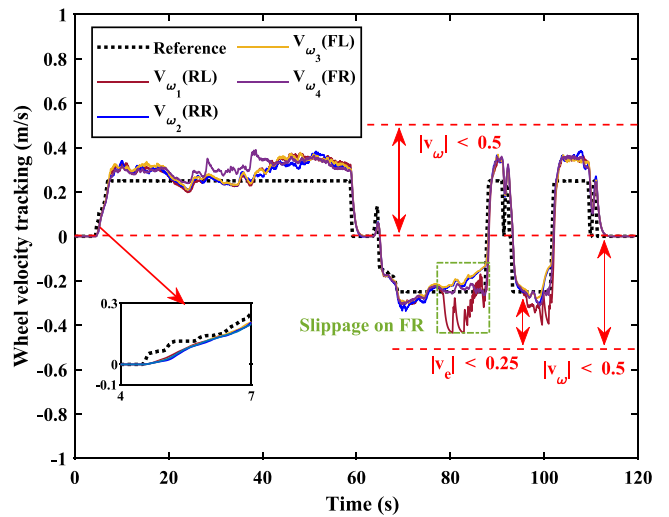

**Fig. 9.** Experiment 1 - Constrained wheel velocity tracking under severe slippage in the RL and FR wheels.

It is well-known that during slippage, the tracking velocity error increases, leading to the generation of excessive control command signals. This, in turn, causes the torque to be required excessively, further destabilizing the soil. However, as shown in Fig. 11, the control valve signals remained within their limits. This behavior is further corroborated by Fig. 12, which illustrates the required torques. For instance, during the severe slippage of the RL wheel, the other motor torques tended to increase excessively to compensate for the disturbance caused by the RL wheel slippage but were effectively constrained within the safety limits, maintaining the whole system stable. Overall, our investigations in this experiment showed that the RL and FR wheels had the highest root mean squared error of tracking due to slippage occurring in them, while the proposed RTOVC maintained stable tracking and adhered to the predefined safety constraints. The RR and FL wheels did not have to compensate for severe slippage and exhibited better accuracy. Hence, during this scenario, the IWD-actuated RR wheel achieved the best velocity tracking, with a mean squared rooted error of approximately 0.07 m/s. This performance surpassed the FL, FR, and RL wheels, which exhibited average errors of 0.08, 0.09, and 0.12 m/s, respectively. In addition, the highest torque effort was associated with the RR wheel, at 280 N· m, while the lowest was recorded for the RL wheel, at 125 N· m.

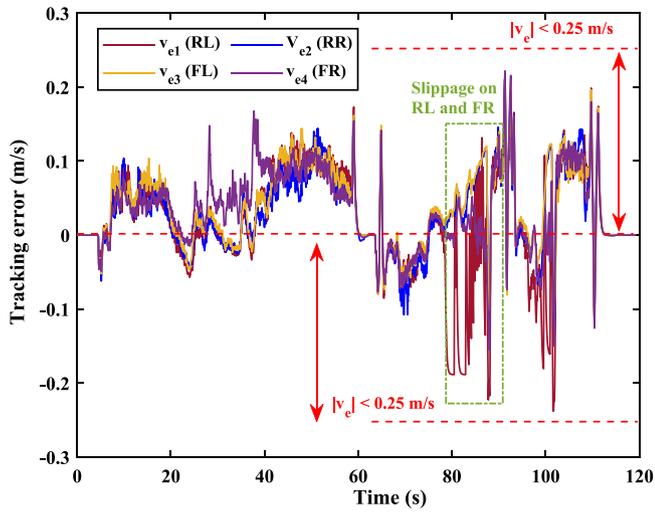

**Fig. 10.** Experiment 1 - Constrained tracking error under severe slippage in the RL and FR wheels.

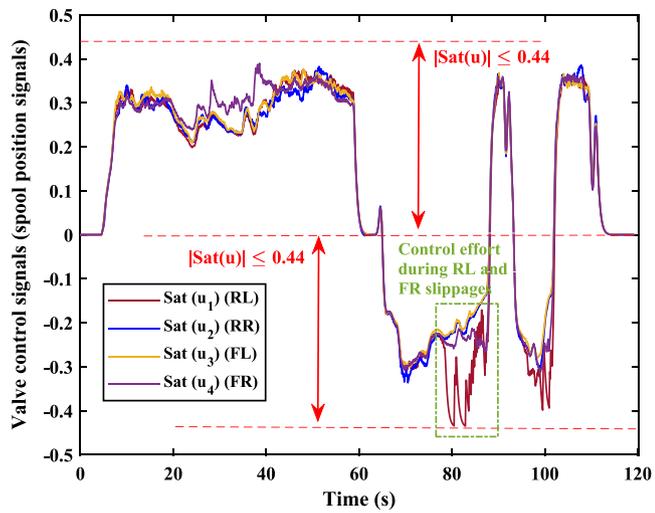

**Fig. 11.** Experiment 1 - Constrained valve control signals under severe slippage in the RL and FR wheels.

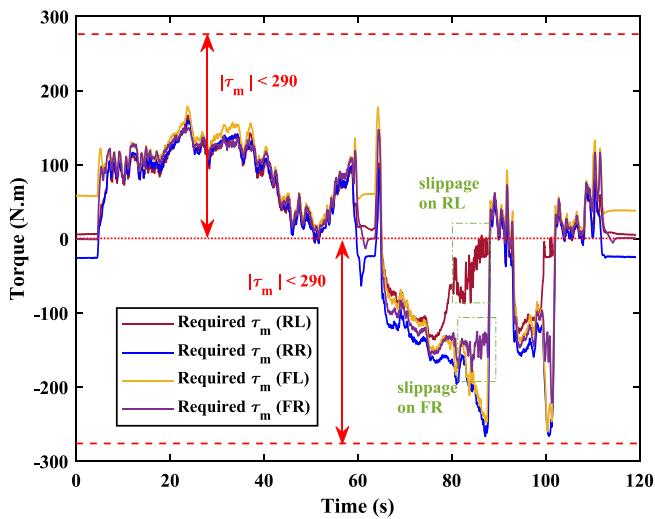

**Fig. 12.** Experiment 1 - Required motors' torques under severe slippage in the RL and FR wheels.

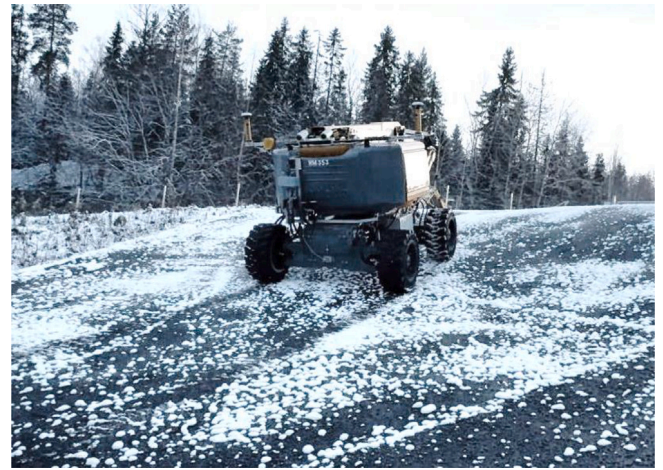

**Fig. 13.** In Experiment 2, the performance of the studied HWMR was evaluated on a slippery, icy, gravel terrain with an approximate slope of 35 degrees. The relevant video is available at: https://youtu.be/AWnTkHOndLU.

### 4.3. Experiment 2: steep ice-covered stony terrain

The large-scale image of the IWD-actuated HWMR on the slippery and icy terrain is shown in Fig. 13. During this execution, all IWD-actuated wheels encountered a high slope and slippery surface. In particular, the FR wheel exhibited severe slippage during forward movement over the steepest section of the road, with a thick layer of ice, occurring between 100 and 115 seconds of the running task. Despite this, the wheels' velocity tracking is shown in Fig. 14, indicating that all wheels, utilizing the RTOVC, tracked the joystick-generated velocity (depicted in black) under severe slippage, ensuring the robustness of control without torque feedback sensing. In addition, similar to Experiment 1, the safety-defined constraints on the actual velocity $|v_o| < 0.5$ m/s and velocity tracking error $|v_e| < 0.25$ m/s which were defined in (31) were met. As observed in Fig. 14, the absolute actual velocities of the IWD-actuated wheels were more than the velocity command due to the increased gravitational forces. The large-scale image in Fig. 14 further illustrates the quick response of the RTOVC and its capability to track changes in the velocities of the IWD-actuated wheels during the slippage of the FR wheel. These details are shown in Fig. 15 more clearly, indicating the actual tracking error signals. The figure clearly demonstrates that all four wheels consistently suffered from slippage, with the FR wheel experiencing more severe slippage. This validates that the control performance remains within the safety-defined error constraints while maintaining robustness in the presence of severe disturbances caused by rough terrain on the studied 6650-kg HWMR. The control valve signals generated by the RTOVC for this scenario are presented in Fig. 16. It generates sufficient valve signals to produce the required torques, consequently tracking the commanded velocity from the joystick. This indicates that all wheels, utilizing the RTOVC, accurately tracked the joystick-generated velocity (depicted in black) even under severe slippage, while maintaining robustness and adhering to safe constraints. As observed during this scenario, the high slope and slippery terrain caused the control signals to be generated excessively, leading to system instability. However, the safety-defined constraints on all four valve signals as (31), $|sat(u)| \leq 0.44$, were adhered to.

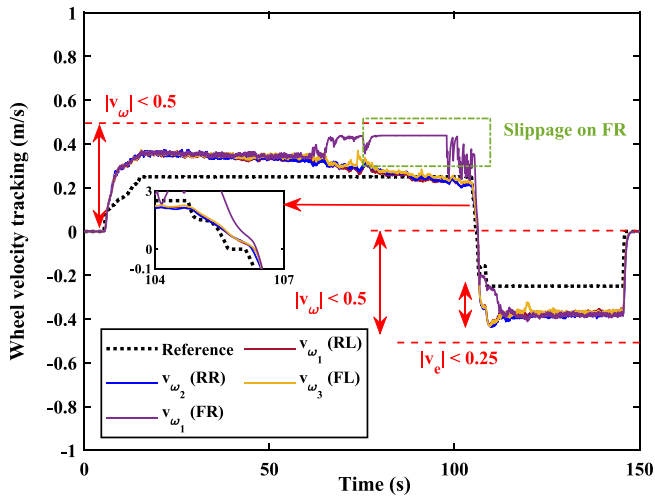

**Fig. 14.** Experiment 2 - Constrained wheel velocity tracking under severe slippage in the FR wheel.

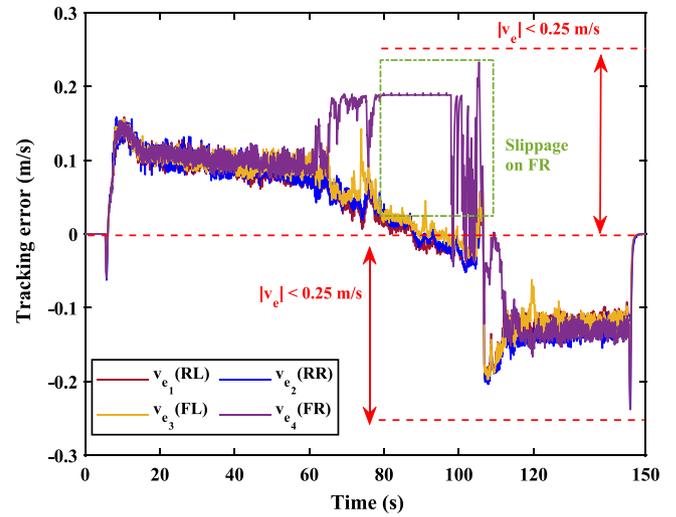

**Fig. 15.** Experiment 2 - Constrained tracking error under severe slippage in the FR wheel.

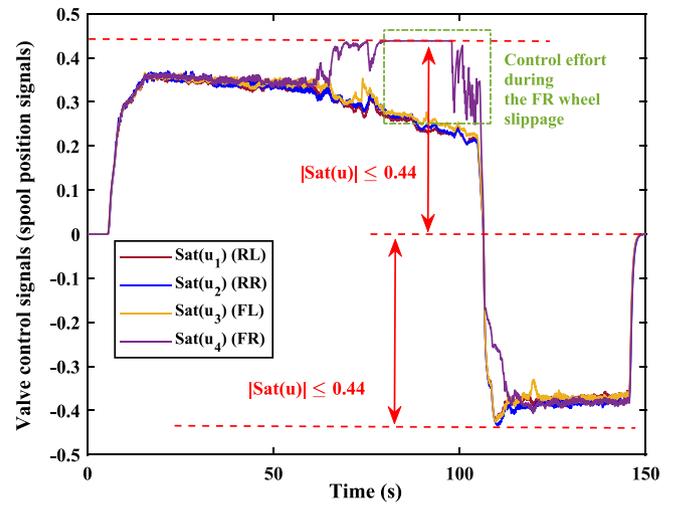

**Fig. 16.** Experiment 2 - Constrained valve control signals under severe slippage in the FR wheel.

Similar to the constrained control valve signals and IWD velocities, the required motor torque never exceeded the safety-defined constraints, $|\tau_m| < 290$, as shown in Fig. 17, even under slippage. As mentioned earlier in Experiment 1, during slippage, the tracking velocity error increases, leading to the generation of excessive control command signals. This, in turn, causes the torque to be generated excessively, further destabilizing the system. However, as shown in Fig. 16, the control valve signals remained within their limits. This behavior is further corroborated by Fig. 17, which illustrates the required torques. This demonstrates the required four motor torques obtained from Eq. (26) to generate sufficient wheel motion for tracking the commanded joystick reference motion while maintaining the stability of the HWMR under slippage on steep icy terrain. For instance, during the severe slippage of the FR wheel, the other motor torques tended to increase excessively to compensate for the disturbance caused by the FR wheel slippage but were effectively constrained within the safety limits, maintaining the whole system stable. Our detailed investigation revealed that, during this scenario, the IWD-actuated RR wheel achieved the best tracking performance, with a root mean squared error of approximately 0.08 m/s. This performance surpassed the FL, RL, and FR wheels, which exhibited average errors of 0.09, 0.09, and 0.13 m/s, respectively. In addition, the highest torque effort was associated with the FL wheel, at 200 N·m, while the lowest was recorded for the FR wheel, at 100 N·m.

As shown in both scenarios, the wheels arbitrarily experienced slippage. However, the RTOVC successfully controlled and maintained the stability of the tracking performance even under these uncertain environmental forces. Table 2 compares the performances of the RTOVC with a model-free robust backstepping-based adaptive control (BAC), as proposed in Cai, Wen, Su, Liu, and Xing (2016), and a model-based robust decentralized-valve-structure control (DVSC), as proposed in Schwarz and Lohmann (2024) for the studied IWD-actuated HWMR in the same condition. The selected tracking criteria—velocity error and torque efforts—were based on the root mean squared velocity tracking error (m/s) and the maximum motor efforts (N.m) during movement to ensure a fairer comparison. Note that unlike the RTOVC, which uses a torque observer, the implementation of the BAC and DVSC schemes utilized torque feedback via the HWMR's pressure sensors without the safety-defined constraints, resorting to the emergency push button when necessary. In addition to uniformly exponential stability, torque observation, and safety-defined constraints, the tracking performances detailed in the table indicate an improvement in controlling the studied four hydraulic-powered IWD-actuated HWMR by employing the RTOVC framework. We also employed the proportional–integral–derivative (PID) controller for the case study, however, we were unable to control the HWMR effectively using the PID controller because it may be adequate only for systems primarily characterized by second-order dynamics (Shahna & Mattila, 2024), while the higher-order-dynamic hydraulic-powered IWD-actuated HWMR exposed severe disturbances due to slipping terrains and uncertainties in hydraulic drive systems.

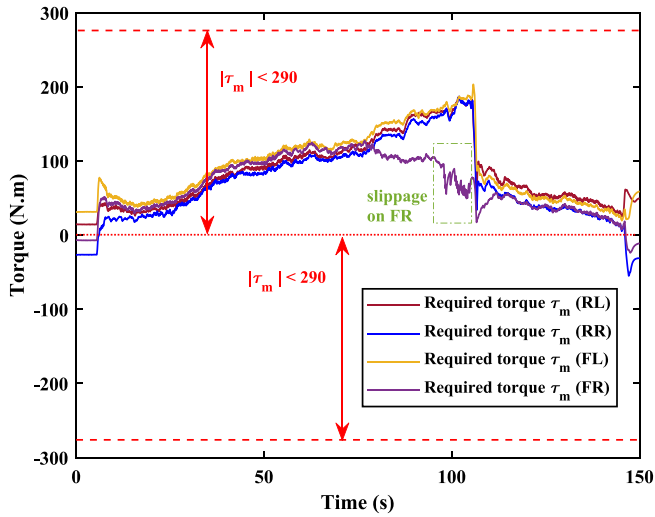

**Fig. 17.** Experiment 2 - Required motors' torques under severe slippage in the FR wheel.

**Table 2**
Performance of the HWMR wheels implemented by the RTOVC, DVSC (Schwarz & Lohmann, 2024), and BAC (Cai et al., 2016) in two experiments under slippage. The recorded errors are based on the root mean squared.

| Exp. | Wheel | Criteria | RTOVC | DVSC (Schwarz & Lohmann, 2024) | BAC (Cai et al., 2016) |
|------|-------|----------|-------|-------------------------------|------------------------|
| 1 | RL | Vel. error / Torque | 0.12/125 | 0.18/220 | 0.25/300 |
| | RR | Vel. error / Torque | 0.07/280 | 0.12/300 | 0.10/310 |
| | FL | Vel. error / Torque | 0.08/275 | 0.14/295 | 0.14/320 |
| | FR | Vel. error / Torque | 0.09/250 | 0.11/290 | 0.18/290 |
| 2 | RL | Vel. error / Torque | 0.09/190 | 0.09/125 | 0.15/145 |
| | RR | Vel. error / Torque | 0.08/190 | 0.16/235 | 0.21/240 |
| | FL | Vel. error / Torque | 0.09/200 | 0.12/280 | 0.30/320 |
| | FR | Vel. error / Torque | 0.13/100 | 0.19/155 | 0.23/205 |

## 5. Conclusion

To address the vital requirement for advancements in designing a strongly stability-guaranteed, safe, and robust control system for hydraulic-powered IWD-actuated HWMR systems in uncertain and time-varying operating conditions, this paper proposed the RTOVC within a logarithmic BLF framework. It decreased the closed-loop dependency associated with fault-prone sensors by obtaining the required torque of each hydraulic-powered IWD-actuated wheel of the distributed HWMR system using a novel robust adaptive BLF-based observer to track the desired velocity. In addition, another robust adaptive BLF-based control was employed for the hydraulic system to manage the valve control signals, generating the same required torque for each wheel. The RTOVC framework including the proposed robust torque observer and robust valve control reduces the risk of system faults by safely constraining key input–output signals—such as the actual velocities, tracking errors, motor/wheel torques, and control valve signals of hydraulic-powered IWD-actuated wheels within logarithmic BLFs. Our study found that a 6500-kg HWMR employing four RTOVC-applied hydraulic-powered IWD-actuated wheels achieved strong robustness and stability of the whole system, even in the presence of safety-defined constraints and disturbances, such as severe wheel slippage on rough terrains.

## CRediT authorship contribution statement

**Mehdi Heydari Shahna:** Writing – original draft, Validation, Software, Methodology, Investigation, Formal analysis, Conceptualization. **Pauli Mustalahti:** Validation, Resources, Investigation, Data curation. **Jouni Mattila:** Writing – review & editing, Supervision, Resources, Funding acquisition.

## Declaration of competing interest

The authors declare that they have no known competing financial interests or personal relationships that could have appeared to influence the work reported in this paper.

## Acknowledgments

This work was supported by the Business Finland Partnership Project, 'Future All-Electric Rough Terrain Autonomous Mobile Manipulators' under Grant No. 2334/31/2022.

## Appendix A. Hydraulic IWD stability

**Proof of Theorem 1.** Define the error of the first adaptation law in Eq. (24) as

$$\bar{\psi}_1 = \hat{\psi}_1 - \psi_1^* \tag{32}$$

where $\psi_1^*$ is an unknown positive constant and can be defined as

$$\psi_1^* = \frac{2}{k_2}(\rho_1 \frac{1}{\bar{a}_1^2} \mu_1^2) \tag{33}$$

where $\rho_1$ and $\mu_1$ are unknown positive constants. $\bar{a}_1$ is a positive constant as $0 < \bar{a}_1 \le a_1$. As a $\psi_1^*$ is positive constant, we use (24) and (32) to provide:

$$\dot{\hat{\psi}}_1 = -k_3 k_4 \bar{\psi}_1 + \frac{1}{2} k_2 k_3 \beta_1^2 - k_3 k_4 \psi_1^* \tag{34}$$

Based on the BLF concept, the logarithmic or fractional barrier function can be designed, tending to infinity as the system state approaches a constraint boundary. Hence, let us introduce a logarithmic BLF as

$$V_1 = \frac{1}{2\bar{a}_1} \log\left(\frac{\alpha_1^2}{Q_1}\right) + \frac{1}{2k_3} \bar{\psi}_1^2 \tag{35}$$

where $Q_1 = \alpha_1^2 - v_e^2$ follows the defined constraint in Eq. (23). If we already choose $|v_e(t_0)| < \alpha_1(t_0)$, the velocity error must satisfy $|v_e| < \alpha_1$, as also observed in (25). Increasing $v_e^2$ and reaching the predefined tracking error bound $\alpha_1^2$ leads to singularities in the system and the execution can systematically stop due to a warning, such as 'Infinity or NaN value encountered', ensuring that the tracking error variable does not exceed the bound $\alpha_1$. Therefore, by taking the derivative of (35) considering (21):

$$\dot{V}_1 = \beta_1 \frac{a_1}{\bar{a}_1} \tau_\omega + \frac{1}{\bar{a}_1} \beta_1 F_1^* + \frac{1}{\bar{a}_1} \beta_1 G_1 + k_3^{-1} \bar{\psi}_1 \dot{\hat{\psi}}_1 \tag{36}$$

Then, from (36) and Assumption 3, we have

$$\dot{V}_1 \le \beta_1 \frac{a_1}{\bar{a}_1} \tau_\omega + \frac{1}{\bar{a}_1} \mu_1 m_1 \mid \beta_1 \mid + \frac{1}{\bar{a}_1} \beta_1 G_1 + k_3^{-1} \bar{\psi}_1 \dot{\hat{\psi}}_1 \tag{37}$$

Now, by utilizing the proposed torque observer $\hat{\tau}_\omega$ provided in (26) instead of $\tau_\omega$ in (37), we have:

$$\begin{aligned} \dot{V}_1 \le -\frac{1}{2}\frac{a_1}{\bar{a}_1}\beta_1 k_1 v_e - \frac{1}{2} k_2 \hat{\psi}_1 \beta_1^2 - k_5 \beta_1^2 G_1^2 + \frac{1}{\bar{a}_1} \mu_1 m_1 \mid \beta_1 \mid \\ + \frac{1}{\bar{a}_1} \beta_1 G_1 + k_3^{-1} \bar{\psi}_1 \dot{\hat{\psi}}_1 \end{aligned} \tag{38}$$

Note that the initial condition of (24) must be $\hat{\psi}_1(t_0) > 0$, guaranteeing $\hat{\psi}_1(t) > 0$ for any $t \ge t_0$. Note $\frac{a_1}{\bar{a}_1} \ge 1$.

**Remark 2.** Based on the Cauchy–Schwarz inequality for any $a$ and $b$ we can say $ab \leq ca^2 + \frac{1}{4c}b^2$, where $c$ is positive.

$k_5$ can be any positive constant. Based on **Remark 2**, we can have

$$\dot{V}_1 \leq -\frac{1}{2}\frac{a_1}{\tilde{a}_1}\beta_1 k_1 v_e - \frac{1}{2}k_2 \tilde{\psi}_1 \beta_1^2 + \frac{1}{4}\rho_1^{-1} m_1^2 + \rho_1(\frac{1}{a_1})^2 \mu_1^2 \beta_1^2 \\ - k_5 \beta_1^2 G_1^2 + k_5 \beta_1^2 G_1^2 + \frac{1}{k_5 \tilde{a}_1^2} + k_3^{-1} \tilde{\psi}_1 \dot{\psi}_1 \tag{39}$$

where $\rho_1$ is any positive constant. Thus, from (33) and (39), we obtain:

$$\dot{V}_1 \leq -\frac{1}{2}\frac{a_1}{\tilde{a}_1}\beta_1 k_1 v_e - \frac{1}{2}k_2 \tilde{\psi}_1 \beta_1^2 + \frac{1}{4}\rho_1^{-1} m_1^2 + \frac{1}{2}k_2 \psi_1^* \beta_1^2 \\ + k_3^{-1} \tilde{\psi}_1 \dot{\psi}_1 \tag{40}$$

Inserting (34) into (40), we obtain:

$$\dot{V}_1 \leq -\frac{1}{2}\frac{a_1}{\tilde{a}_1}\beta_1 k_1 v_e - \frac{1}{2}k_2(\tilde{\psi}_1 - \psi_1^*)\beta_1^2 + \frac{1}{4}\rho_1^{-1} m_1^2 - k_4 \tilde{\psi}_1^2 \\ + \frac{1}{2}k_2 \tilde{\psi}_1 \hat{\tau}_m - k_4 \psi_1^* \tilde{\psi}_1 \tag{41}$$

Recall that $\beta_1 = \frac{v_e}{Q_1}$ and $\frac{a_1}{\tilde{a}_1} \geq 1$. Thus:

$$\dot{V}_1 \leq -\frac{1}{2}k_1 \frac{v_e^2}{Q_1} + \frac{1}{4}\rho_1^{-1} m_1^2 - \frac{1}{2}k_4 \tilde{\psi}_1^2 - \frac{1}{2}k_4(\tilde{\psi}_1 - \psi_1^*)^2 \\ - k_4 \psi_1^*(\tilde{\psi}_1 - \psi_1^*) \tag{42}$$

Note that the initial condition of (24) must be $\hat{\psi}_1(t_0) > 0$, guaranteeing $\hat{\psi}_1(t) > 0$ for any $t \geq t_0$. Thus, by knowing $\hat{\psi}_1$ is always positive, we can Simplify (42), as:

$$\dot{V}_1 \leq -\frac{1}{2}k_1 \frac{v_e^2}{Q_1} + \frac{1}{4}\rho_1^{-1} m_1^2 - \frac{1}{2}k_4 \tilde{\psi}_1^2 + \frac{1}{2}k_4 \psi_1^{*2} \tag{43}$$

According to Ren, Ge, Tee, and Lee (2010) and knowing $Q_1 = \alpha_1^2 - v_e^2$, if $|v_e| < \alpha_1$, we have:

$$\log\left(\frac{\alpha_1^2}{Q_1}\right) < \frac{v_e^2}{Q_1} \tag{44}$$

Thus, from (43) and (44), we have:

$$\dot{V}_1 \leq -\frac{1}{2}k_1 \log\left(\frac{\alpha_1^2}{Q_1}\right) + \frac{1}{4}\rho_1^{-1} m_1^2 - \frac{1}{2}k_4 \tilde{\psi}_1^2 + \frac{1}{2}k_4 \psi_1^{*2} \tag{45}$$

or simply from (35) and (45), we can say

$$\dot{V}_1 \leq -\Omega_{11} V_1 + \frac{1}{4}\rho_1^{-1} m_1^2 + \Omega_{12} \tag{46}$$

where

$$\Omega_{11} = \min\left[\ \tilde{a}_1 k_1, \quad k_3 k_4\ \right], \quad \Omega_{12} = \frac{1}{2}k_4 \psi_1^{*2} \tag{47}$$

Note that $\psi_1^*$ is based on the bound of uncertainties and external disturbances $\mu_1$ due to the modeling error and wheel slip (see (33)). Let us define the error of the second adaptation law, as:

$$\bar{\psi}_2 = \hat{\psi}_2 - \psi_2^* \tag{48}$$

where $\psi_2^*$ is an unknown positive constant and is introduced as

$$\psi_2^* = \frac{2}{k_7}(\rho_2 \frac{1}{\tilde{a}_2^2} \mu_2^2) \tag{49}$$

$\rho_2$ and $\mu_2$ are unknown positive constants. As $\psi_2^*$ is a positive constant, we use (28) and (48):

$$\dot{\hat{\psi}}_2 = -k_8 k_9 \bar{\psi}_2 + \frac{1}{2}k_7 k_8 \beta_2^2 - k_8 k_9 \psi_2^* \tag{50}$$

Based on the BLF concept, the logarithmic or fractional barrier function can be designed, tending to infinity as the system state approaches a constraint boundary. As observed in (29), if $|\hat{\tau}_m(t_0)| < \tau_{max}(t_0)$, the required torque must satisfy $|\hat{\tau}_m| < \tau_{max}$. Increasing $\hat{\tau}_m^2$ and reaching the predefined bound $\tau_{max}^2$ leads to singularities in the system. For instance, in many programming platforms, execution can stop due to a warning, such as 'Infinity or NaN value encountered,' when the computed value exceeds the system's numerical limits. It can prevent the issue of excessive motor torque generation, a common problem in velocity-based control systems, particularly during wheel slippage. Hence, let us introduce another logarithmic BLF as

$$V_2 = \frac{1}{2\tilde{a}_2}\log\left(\frac{\tau_{max}^2}{Q_2}\right) + \frac{1}{2k_8}\bar{\psi}_2^2 \tag{51}$$

where $Q_2 = \tau_{max}^2 - \hat{\tau}_m^2$. Recall that $\beta_2 = \frac{\hat{\tau}_m}{Q_2}$. Thus, by taking the derivative of (51) and considering (17):

$$\dot{V}_2 = \beta_2 \frac{a_2}{\tilde{a}_2} u + \frac{1}{\tilde{a}_2}\beta_2 F_2 + k_8^{-1}\bar{\psi}_2 \dot{\psi}_2 \tag{52}$$

Then, from (18) in Assumption 2, and (52):

$$\dot{V}_2 \leq \beta_2 \frac{a_2}{\tilde{a}_2} u + \frac{1}{\tilde{a}_2}\mu_2 m_2 \mid \beta_2 \mid + k_8^{-1}\bar{\psi}_2 \dot{\psi}_2 \tag{53}$$

Now, by inserting the proposed valve control signal provided in (30) into (53), we have:

$$\dot{V}_2 \leq -\frac{1}{2}\beta_2 k_6 \hat{\tau}_m - \frac{1}{2}k_7 \hat{\psi}_2 \beta_2^2 + \frac{1}{\tilde{a}_2}\mu_2 m_2 \mid \beta_2 \mid + k_8^{-1}\bar{\psi}_2 \dot{\psi}_2 \tag{54}$$

Note that the initial condition of (28) must be $\hat{\psi}_2(t_0) > 0$, guaranteeing $\hat{\psi}_2(t) > 0$ for any $t \geq t_0$. Consider that $\frac{a_2}{\tilde{a}_2} \geq 1$. Consider $\rho_2$ is any positive constant. Based on **Remark 2**, we have

$$\dot{V}_2 \leq -\frac{1}{2}\frac{a_2}{\tilde{a}_2}\beta_2 k_6 \hat{\tau}_m - \frac{1}{2}k_7 \hat{\psi}_2 \beta_2^2 + \frac{1}{4}\rho_2^{-1} m_2^2 + \rho_2(\frac{1}{\tilde{a}_2})^2 \mu_2^2 \beta_2^2 \\ + k_8^{-1}\bar{\psi}_2 \dot{\psi}_2 \tag{55}$$

Thus, from (49) and (55), we obtain:

$$\dot{V}_2 \leq -\frac{1}{2}\frac{a_2}{\tilde{a}_2}\beta_2 k_6 \hat{\tau}_m - \frac{1}{2}k_7 \hat{\psi}_2 \beta_2^2 + \frac{1}{4}\rho_2^{-1} m_2^2 + \frac{1}{2}k_7 \psi_2^* \beta_2^2 + k_8^{-1}\bar{\psi}_2 \dot{\psi}_2 \tag{56}$$

Inserting (50) into (56), we obtain:

$$\dot{V}_2 \leq -\frac{1}{2}\frac{a_2}{\tilde{a}_2}\beta_2 k_6 \hat{\tau}_m - \frac{1}{2}k_7(\hat{\psi}_2 - \psi_2^*)\beta_2^2 + \frac{1}{4}\rho_2^{-1} m_2^2 - k_9 \bar{\psi}_2^2 \\ + \frac{1}{2}k_9 \bar{\psi}_2 \beta_2^2 - k_9 \psi_2^* \bar{\psi}_2 \tag{57}$$

Recall that $\beta_2 = \frac{\hat{\tau}_m}{Q_2}$ and $\frac{a_2}{\tilde{a}_2} \geq 1$. Thus:

$$\dot{V}_2 \leq -\frac{1}{2}k_6 \frac{\hat{\tau}_m^2}{Q_2} + \frac{1}{4}\rho_2^{-1} m_2^2 - \frac{1}{2}k_9 \bar{\psi}_2^2 - \frac{1}{2}k_9(\hat{\psi}_2 - \psi_2^*)^2 \\ - k_9 \psi_2^*(\hat{\psi}_2 - \psi_2^*) \tag{58}$$

Thus, by knowing $\hat{\psi}_2$ is always positive, we can Simplify (58), as:

$$\dot{V}_2 \leq -\frac{1}{2}k_6 \frac{\hat{\tau}_m^2}{Q_2} + \frac{1}{4}\rho_2^{-1} m_2^2 - \frac{1}{2}k_9 \bar{\psi}_2^2 + \frac{1}{2}k_9 \psi_2^{*2} \tag{59}$$

According to Ren et al. (2010) and knowing $Q_2 = \tau_{max}^2 - \hat{\tau}_m^2$, if $|\hat{\tau}_m| < \tau_{max}^2$, we have:

$$\log\left(\frac{\tau_{max}^2}{Q_2}\right) < \frac{\hat{\tau}_m^2}{Q_2} \tag{60}$$

Thus, from (59) and (60), we have:

$$\dot{V}_2 \leq -\frac{1}{2}k_6 \log\left(\frac{\tau_{max}^2}{Q_2}\right) + \frac{1}{4}\rho_2^{-1} m_2^2 + \frac{1}{4}\rho_2^{-1} - \frac{1}{2}k_9 \bar{\psi}_2^2 + \frac{1}{2}k_9 \psi_2^{*2} \tag{61}$$

or simply from (61), we can say:

$$\dot{V}_2 \leq -\Omega_{21} V_2 + \frac{1}{4}\rho_2^{-1} m_2^2 + \Omega_{22} \tag{62}$$

where

$$\Omega_{21} = \min\left[\ \tilde{a}_2 k_6, \quad k_8 k_9\ \right], \quad \Omega_{22} = \frac{1}{2}k_9 \psi_2^{*2} \tag{63}$$

$\psi_2^*$ is based on the bound of uncertainties and external disturbances $\mu_2$ is the hydraulic servo mechanism (see (49)). Now, consider one hydraulic-powered IWD-actuated wheel, which utilizes the proposed RTOVC framework, including the torque control outlined in (26) and the valve control signal described in (30). Let us define a quadratic function, which is the sum of the quadratic functions (35) and (51), as follows:

$$\bar{V} = V_1 + V_2 \\ = \frac{1}{2\tilde{a}_1}\log\left(\frac{\alpha_1^2}{Q_1}\right) + \frac{1}{2\tilde{a}_2}\log\left(\frac{\tau_{max}^2}{Q_2}\right) + \frac{1}{2k_3}\tilde{\psi}_1^2 + \frac{1}{2k_8}\bar{\psi}_2^2 \tag{64}$$

For simplification, assume $\alpha_2 = \tau_{max}$. Now, let us define the following notation for $i = 1, 2$:

$$\Theta_i = \log\left(\frac{\alpha_i^2}{Q_i}\right) \tag{65}$$

Based on (64), we have:

$$\bar{V} = \frac{1}{2a_1}\Theta_1 + \frac{1}{2a_2}\Theta_2 + \frac{1}{2k_3}\bar{\psi}_1^2 + \frac{1}{2k_8}\bar{\psi}_2^2 \tag{66}$$

The derivative of (66), based on (46) and (62), can be expressed, as:

$$\dot{\bar{V}} = \dot{V}_1 + \dot{V}_2 \leq -\Omega\bar{V} + \frac{1}{4}\sum_{i=1}^{2}\rho_i^{-1}m_i^2 + \bar{\Omega} \tag{67}$$

where

$$\Omega = \min\{\Omega_{11}, \Omega_{21}\} = \min\{\bar{a}_1 k_1, \bar{a}_2 k_6, k_3 k_4, k_8 k_9\}$$
$$\bar{\Omega} = \Omega_{12} + \Omega_{22} = \frac{1}{2}k_4\psi_1^{*2} + \frac{1}{2}k_9\psi_2^{*2} \tag{68}$$

Therefore:

$$\bar{V} \leq \bar{V}(t_0)\, e^{-\{\Omega(t-t_0)\}} + \bar{\Omega}\,\Omega^{-1} + \frac{1}{4}\sum_{i=1}^{2}\rho_i^{-1}\int_{t_0}^{t}e^{-\Omega(t-T)}m_i^2\;dT \tag{69}$$

From (65) and (66), we can say for $i = 1, 2$:

$$\Theta_i^2 \leq 2\bar{a}_i\bar{V}(t_0)\, e^{-\{\Omega(t-t_0)\}} + 2\bar{a}_i\,\bar{\Omega}\,\Omega^{-1} + \frac{1}{2}\bar{a}_i\sum_{i=1}^{2}\rho_i^{-1}\int_{t_0}^{t}e^{\{-\Omega(t-T)\}}m_i^2\;dT \tag{70}$$

A continuous operator can be suggested as:

$$Z(\iota) = \frac{0.5}{\Omega - \iota} > 0 \tag{71}$$

where $\iota \in [0, \Omega)$. Note $Z(0) = \frac{1}{2\Omega} \leq Z(\iota)$. We are able to find a sufficient small parameter $\bar{\iota} \in \iota$ that satisfies the following condition:

$$0 < \overset{*}{Z} = Z(\bar{\iota}) = \frac{0.5}{\Omega - \bar{\iota}} < 1 \tag{72}$$

By multiplying $e^{\bar{\iota}(t-t_0)}$, we reach:

$$\Theta_i e^{\bar{\iota}(t-t_0)} \leq 2\bar{a}_i\bar{V}(t_0)e^{-(\Omega-\bar{\iota})(t-t_0)} + 2\bar{a}_i\bar{\Omega}\Omega^{-1}e^{\bar{\iota}(t-t_0)}$$
$$+ \frac{1}{2}\bar{a}_i\sum_{i=1}^{2}\rho_i^{-1}\int_{t_0}^{t}e^{-\Omega(t-T)+\bar{\iota}(t-t_0)}m_i^2\;dT \tag{73}$$

We can omit the reducing element that existed on the right-hand side, as:

$$\Theta_i e^{\bar{\iota}(t-t_0)} \leq 2\bar{a}_i\bar{V}(t_0) + 2\bar{a}_i\bar{\Omega}\Omega^{-1}e^{\bar{\iota}(t-t_0)}$$
$$+ \frac{1}{2}\bar{a}_i\sum_{i=1}^{2}\rho_i^{-1}\int_{t_0}^{t}e^{-\Omega(t-T)+\bar{\iota}(t-t_0)}m_i^2\;dT \tag{74}$$

Using this approach, we can describe functions $E_0$ and $E_1$, which are continuous:

$$E_0 = \Theta_i^2 e^{\bar{\iota}(w-t_0)}$$
$$E_1 = \sup_{w\in[t_0,t]}\left[\sum_{i=1}^{2}\rho_i^{-1}(m_i^2)e^{\bar{\iota}(t-t_0)}\right] \tag{75}$$

Next, by considering (74) and (75), and knowing $\bar{\iota} \geq 0$, we obtain:

$$E_0 \leq 2\bar{a}_i\bar{V}(t_0) + 2\bar{a}_i\bar{\Omega}\Omega^{-1} + \frac{0.5\bar{a}_i}{\Omega - \bar{\iota}}E_1 \tag{76}$$

By defining $\bar{E} = \max(E_0, E_1)$, we can have:

$$E_0 \leq 2\bar{a}_i\bar{V}(t_0) + 2\bar{a}_i\bar{\Omega}\Omega^{-1}e^{\bar{\iota}(t-t_0)} + \overset{*}{Z}\bar{a}_i\bar{E} \tag{77}$$

It is always possible to ensure the existence of $\bar{\iota}$ and $\iota^*$, where $\overset{*}{Z}$ sufficiently small in (72) and $\Omega > \iota^* > \bar{\iota}$, to satisfy the following condition (Shahna, Bahari, & Mattila, 2024a):

$$\overset{*}{Z} = Z(\iota^*) > \overset{*}{Z}, \quad 0 < \overset{*}{Z} < 1, \quad \overset{*}{Z}\bar{a}_i\bar{E} \leq \overset{*}{Z}\bar{a}_i E_0 \tag{78}$$

Inserting (78) into (77), we arrive at:

$$E_0 \leq 2\bar{a}_i\bar{V}(t_0) + 2\bar{a}_i\bar{\Omega}\Omega^{-1}e^{\bar{\iota}(t-t_0)} + \overset{*}{Z}E_0 \tag{79}$$

Afterward, we have:

$$E_0 \leq \frac{2\bar{a}_i\bar{V}(t_0) + 2\bar{a}_i\bar{\Omega}\Omega^{-1}e^{\bar{\iota}(t-t_0)}}{1 - \overset{*}{Z}} \tag{80}$$

Concerning the definition of $E_0$ in (75), we achieve:

$$\Theta_i^2 \leq \frac{2\bar{a}_i\bar{V}(t_0)\, e^{-\bar{\iota}(t-t_0)} + 2\bar{a}_i\bar{\Omega}\Omega^{-1}}{1 - \overset{*}{Z}} \tag{81}$$

Based on Minkowski's inequality, we reach:

$$\|\Theta_i\| \leq \sqrt{\frac{2\bar{a}_i\bar{V}(t_0)}{1 - \overset{*}{Z}}}\,e^{-\frac{\bar{\iota}}{2}(t-t_0)} + \sqrt{\frac{2\bar{a}_i\bar{\Omega}\Omega^{-1}}{1 - \overset{*}{Z}}} \tag{82}$$

**Definition 1.** According to Corless and Leitmann (1993), Shahna, Bahari, and Mattila (2024b), Shahna, Kolagar, and Mattila (2024), the system state $\Theta_i$ for $t \geq t_0$ is uniformly exponentially stable if

$$\|\Theta_i\| \leq \bar{c}e^{-\bar{b}(t-t_0)}\|\Theta_i(t_0)\| + \zeta \tag{83}$$

where $\bar{c}$, $\zeta$, and $\bar{b} \in \mathbb{R}^+$ are positive constants. Note $\bar{b}$ is the exponential rate of the system convergence and $\Theta_i(t_0)$ is the initial state value of Eq. (65). As $t \to \infty$, a stable region $C$ with a radius $\zeta$ can take shape as

$$C := \{\Theta_i \mid \|\Theta_i\| \leq \zeta\} \tag{84}$$

Thus, based on Definition 1 and Eq. (82), Theorem 1 is proved. Note that, based on Shahna et al. (2024b) and Eq. (83), the convergence rate $\bar{b}$ and the radius of the stable region $\zeta$ depend on the intensity of uncertainties and disturbances. Hence, to balance robustness and real-time responsiveness, the proposed control parameters should be carefully tuned according to the intensity of uncertainties and disturbances.

## Appendix B. IWD-actuated HWMR stability

**Proof of Theorem 2.** If we assign the quadratic function provided in (64) for all four independent IWD mechanisms in a four-wheel HWMR, we obtain a quadratic function as

$$\bar{V} = \sum_{j=1}^{4}\bar{V}_j = \frac{1}{2}\sum_{j=1}^{4}\frac{1}{\bar{a}_{1,j}}\Theta_{1,j} + \frac{1}{\bar{a}_{2,j}}\Theta_{2,j} + \frac{1}{k_{3,j}}\bar{\psi}_{1,j}^2 + \frac{1}{k_{8,j}}\bar{\psi}_{2,j}^2 \tag{85}$$

where

$$\Theta_{i,j} = \log\left(\frac{\alpha_{i,j}^2}{Q_{i,j}}\right) \tag{86}$$

$j$ assigns the wheel number for the quadratic function provided in (64). By differentiating (85) and based on (67), we have

$$\dot{\bar{V}} \leq -\bar{\Omega}\bar{V} + \frac{1}{2}\sum_{j=1}^{4}\sum_{i=1}^{2}\rho_{i,j}^{-1}m_{i,j}^2 + \bar{\bar{\Omega}} \tag{87}$$

where $\bar{\Omega} = \min\{\Omega_j \mid j = 1, \ldots, 4\}$. $\Omega_j$ is the very $\Omega$ provided in (68) for $j$th wheel. Similarly $\bar{\bar{\Omega}} = \sum_{j=1}^{4}\bar{\Omega}_j$. By performing the same mathematical manipulations as the equations provided from (69) to (86), we obtain:

$$\|\Theta_{i,j}\| \leq \sqrt{\frac{2\bar{a}_{i,j}\bar{V}(t_0)}{1 - \overset{*}{Z}}}\,e^{-\frac{\bar{\iota}}{2}(t-t_0)} + \sqrt{\frac{2\bar{a}_{i,j}\bar{\bar{\Omega}}\bar{\Omega}^{-1}}{1 - \overset{*}{Z}}} \tag{88}$$

Thus, based on Shahna et al. (2024b), Shahna et al. (2024a), Shahna, Kolagar, and Mattila (2024), and Definition 1, Theorem 2 is proved. Thus, the margins and convergence rate of stability in RTOVC-employed HWMRs were comprehensively analyzed in Appendices A and B. Both the wheel motion dynamics in Eq. (2) and the hydraulic servomechanism system in Eq. (8) include uncertain parameters and functional terms, which mathematically model real-world uncertainties, such as wheel slippage and uneven torque distribution on rough terrain.

## Appendix C. A guide on parameter tuning

For each hydraulic-powered IWD-actuated wheel, the following steps are repeated individually:

- Step (1) Define safety constraints for the velocity tracking error of the wheel $\alpha_1$, the actual velocity of the wheel $\epsilon_1$, the hydraulic motor torque $\tau_{max}$, and valve control signal $u_{max}$.

- Step (2) As we assumed in (38) and (39), $k_5 \geq a_1^{-1} = \frac{J_\omega}{r}$. Thus, $k_5$ is chosen large enough.

- Step (3) For each wheel, separately, $k_1$ and $k_6$ are carefully selected, considering the trade-off between robustness and the tracking system's responsiveness. Increasing these values will enhance the wheel system's robustness against disturbances, wheel slippage, and the intensity of uncertainties in the hydraulic-powered IWD while reducing tracking accuracy. As the slippage potential of the field and uncertainties are unknown, the chosen $k_1$ and $k_6$ are adjusted by increasing them slightly until the actual velocity of the hydraulic IWD-actuated wheel tracks the desired velocity, meaning that they compensate for the disturbance and uncertainties.

- Step (4) When the tracking error converges to zero with the parameters chosen in Steps 1 and 2, adjust the other parameters ($k_2$, $k_3$, $k_4$, $k_7$, $k_8$, and $k_9$) as needed. Interestingly, changing these parameters does not visibly affect the overall control performance. However, based on our experience, their adjustment slightly influences the convergence rate, reducing the required step times for the RTOVC.

- Step (5) Go back to Step (2) if the tracking performance does not reach a desired level.

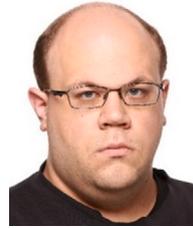

**Pauli Mustalahti** received his M.Sc. degree in engineering from the Tampere University of Technology in 2016 and his D.Sc. (Tech.) degree in Automation Science and Engineering from Tampere University in 2023. He is a researcher in the Unit of Automation Technology and Mechanical Engineering, Tampere University, Tampere, Finland. His research interests include nonlinear model-based control of robotic manipulators.

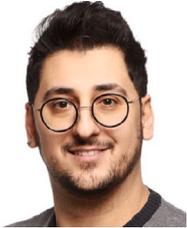

**Mehdi Heydari Shahna** earned a B.Sc. in electrical engineering from Razi University, Kermanshah, Iran, in 2015 and an M.Sc. in control engineering at Shahid Beheshti University, Tehran, Iran, in 2018. Since December 2022, he has been pursuing his doctoral degree in automation technology and mechanical engineering at Tampere University, Tampere, Finland. His research interests encompass robust control, robotics, fault-tolerant algorithms, and system stability.

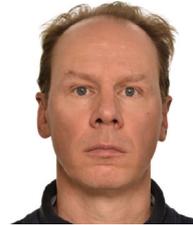

**Jouni Mattila** received an M.Sc. and Ph.D. in automation engineering from Tampere University of Technology, Tampere, Finland, in 1995 and 2000, respectively. He is currently a professor of machine automation in the Automation Technology and Mechanical Engineering Unit at Tampere University. His research interests include machine automation, nonlinear-model-based control of robotic manipulators, and energy-efficient control of heavy-duty mobile manipulators.